\documentclass[showpacs,preprintnumbers,twocolumn,
amsmath,amssymb,groupedaddress,superscriptaddress]{revtex4}

\usepackage{graphicx}
\usepackage{dcolumn}
\usepackage{bm}

\begin{document}

\title{Temperature dependence of the symmetry energy and neutron
skins\\
in Ni, Sn, and Pb isotopic chains}

\author{A. N. Antonov}
\affiliation{Institute for Nuclear Research and Nuclear Energy,
Bulgarian Academy of Sciences, Sofia 1784, Bulgaria}

\author{D. N. Kadrev}
\affiliation{Institute for Nuclear Research and Nuclear Energy,
Bulgarian Academy of Sciences, Sofia 1784, Bulgaria}

\author{M. K. Gaidarov}
\affiliation{Institute for Nuclear Research and Nuclear Energy,
Bulgarian Academy of Sciences, Sofia 1784, Bulgaria}

\author{P. Sarriguren}
\affiliation{Instituto de Estructura de la Materia, IEM-CSIC,
Serrano 123, E-28006 Madrid, Spain}

\author{E. Moya de Guerra}
\affiliation{Grupo de F\'{i}sica Nuclear, Departamento de
F\'{i}sica At\'{o}mica, Molecular y Nuclear,\\ Facultad de
Ciencias F\'{i}sicas, Universidad
Complutense de Madrid, E-28040 Madrid, Spain}


\begin{abstract}
The temperature dependence of the symmetry energy for isotopic
chains of even-even Ni, Sn, and Pb nuclei is investigated in the
framework of the local density approximation (LDA). The Skyrme
energy density functional with two Skyrme-class effective
interactions, SkM* and SLy4, is used in the calculations. The
temperature-dependent proton and neutron densities are calculated
through the HFBTHO code that solves the nuclear
Skyrme-Hartree-Fock-Bogoliubov problem by using the cylindrical
transformed deformed harmonic-oscillator basis. In addition, two
other density distributions of $^{208}$Pb, namely the Fermi-type
density determined within the extended Thomas-Fermi (TF) method
and symmetrized-Fermi local density obtained within the rigorous
density functional approach, are used. The kinetic energy
densities are calculated either by the HFBTHO code or, for a
comparison, by the extended TF method up to second order in
temperature (with $T^{2}$ term). Alternative ways to calculate the
symmetry energy coefficient within the LDA are proposed. The
results for the thermal evolution of the symmetry energy
coefficient in the interval $T=0-4$ MeV show that its values
decrease with temperature. The temperature dependence of the
neutron and proton root-mean-square radii and corresponding
neutron skin thickness is also investigated, showing that the
effect of temperature leads mainly to a substantial increase of
the neutron radii and skins, especially in the more neutron-rich
nuclei, a feature that may have consequences on astrophysical
processes and neutron stars.
\end{abstract}


\maketitle

\section{Introduction}

In recent years, many studies have been carried out to understand
the density dependence of the nuclear equation of state (EOS) over
a wide range of densities and temperatures (see, e.g.,
Ref.~\cite{Lattimer2007,Li2008} and the topical issue of the
European Physical Journal A on nuclear symmetry energy (NSE)
\cite{NSE2014}). This is needed for a reliable treatment of a large variety of
nuclear and astrophysical phenomena. One very important ingredient
of the EOS from both experimental and theoretical aspects is the
symmetry energy that describes the dependence of the energy per
nucleon on the proton to neutron ratio. It is important to
distinguish between finite nuclei and infinite nuclear matter,
where for the latter, the Coulomb interaction is turned off.
Nuclear matter is characterized by its energy per particle as a
function of density and other thermodynamic quantities (e.g.,
temperature). At the same time, within, e.g., the local-density
approximation (LDA) \cite{Agrawal2014,Samaddar2007,Samaddar2008, De2012}
or coherent density fluctuation model (CDFM)
\cite{Gaidarov2011,Gaidarov2012,Gaidarov2014}, one can use the EOS
of asymmetric nuclear matter (ANM) to obtain information on finite
systems.

The nuclear symmetry energy, as a fundamental quantity in nuclear
physics and astrophysics, represents a measure of the energy gain
in converting isospin asymmetric nuclear matter to a symmetric
system. Its value depends on the density $\rho$ and temperature
$T$. Experimentally, the nuclear symmetry energy is not a directly
measurable quantity and is extracted indirectly from observables
that are related to it (e.g., \cite{Shetty2010,Kowalski2007}). The
need of information for the symmetry energy in finite nuclei
(including the one theoretically obtained) is a major issue
because it allows one to constrain the bulk and surface properties
of the nuclear energy-density functionals (EDFs) quite
effectively. More information on the nuclear symmetry energy is
still required for understanding the structures of nuclei far away
from the $\beta$-stability line, heavy-ion collisions, supernova
explosions, and neutron star properties. As can be seen, e.g., in
Refs.~\cite{Li2006,Piekarewicz2009,Vidana2009,Sammarruca2009,Dan2010},
an increasingly wide range of theoretical ideas are being proposed
on the density dependence of the symmetry energy as well as on
some associated nuclear characteristics. In the last years, the
temperature dependence of single-particle properties in nuclear
and neutron matter was also broadly investigated including studies
in finite systems, as well (e.g.,
Refs.~\cite{Moustakidis2007,Sammarruca2008,Agrawal2014,De2014,Zhang2014,Lee2010,BLi2006,
Mekjian2005,Xu2007}).

The thermal behavior of the symmetry energy has a role in changing
the location of the nuclear drip lines as nuclei warm up.
Also, it is of fundamental importance for the liquid-gas phase
transition of asymmetric nuclear matter, the dynamical evolution
mechanisms of massive stars and the supernova explosion
\cite{Baron85}. Since the density derivative of the symmetry
coefficient reflects the pressure difference on the neutrons and
protons and is thus one of the determinants in fixing the neutron
skin of nuclei, the nature and stability of phases within a warm
neutron star, its crustal composition or its thickness
\cite{Steiner2008} would be strongly influenced by the temperature
dependence of the symmetry energy.

The problem of accurate treatment of the thermodynamical
properties of hot finite nuclei is still challenging. Since the
pioneering work of Brack and Quentin \cite{Brack74} on thermal
Hartree-Fock (HF) calculations various methods have been developed
to study the dynamical evolution of such excited systems. Among
them we note semiclassical approaches based on the microscopic
Skyrme-HF formalism \cite{Brack85} and Thomas-Fermi (TF) approximation
\cite{Suraud87} with inclusion of the continuum effects in HF
calculations at finite temperature \cite{Bonche}. Further, refined
Thomas-Fermi description of hot nuclei was reported in
Ref.~\cite{De96}. The extended Thomas-Fermi (ETF) model proposed
by Brack in Ref.~\cite{Brack84} through inclusion of second-order
gradient corrections to the TF density functionals showed their
decisive role in obtaining an excellent agreement with HF results.
More recently, Hartree-Fock-Bogoliubov (HFB) models
\cite{Khan2007,Sandulescu2004,Monrozeau2007} and finite
temperature HF+BCS approximation with zero-range Skyrme forces
\cite{Yuksel2014} have been developed. A relativistic TF
approximation with different relativistic mean-field (RMF) nuclear
interactions has been also explored to extract the symmetry energy
coefficient for several representative nuclei and to study its
temperature dependence \cite{Zhang2014}.

A sensitive probe of the nuclear symmetry energy is the
neutron-skin thickness of nuclei (see, for example,
Ref.~\cite{Sarriguren2007} and references therein). The latter is
commonly defined in terms of the difference between the neutron
and proton root-mean-square (rms) radii and is found to be closely
related to the density dependence of the NSE, with the EOS of pure
neutron matter and properties of neutron stars
\cite{Tondeur84,Horowitz2001,Typel2001,Yoshida2004,Warda2009,Fattoyev2012,Erler2013,Lee2013}.
It is also related to a number of observables in finite nuclei,
including the NSE (see, e.g.,
\cite{Agrawal2014,Brown2000,Furnstahl2002,Reinhard2010,RocaMaza2011,Kortelainen2013,Mondal2016,
Gaidarov2011,Gaidarov2012,Antonov2016,Warda2010,Centelles2010,Dan2003,Dan2004,Dan2014,Dan2009,Dan2011,Tsang2012,Tsang2009,Ono2004,Dan2006,
Diep2007}), although its precise measurement is difficult to be
done. As examples, in Ref.~\cite{Yuksel2014} Y\"{u}ksel  {\it et
al.} have analyzed the temperature dependence of the nuclear radii
for $^{120}$Sn nucleus and neutron skin as a function of $N/Z$
value for tin isotopic chain within the finite temperature HF+BCS
framework using Skyrme interactions. The same nuclear
characteristics have been computed within the relativistic TF
approximation for $^{56}$Fe and $^{208}$Pb nuclei in
Ref.~\cite{Zhang2014}, where both neutron and proton rms radii are
found to increase significantly with increasing $T$, comparable to
those shown in Ref.~\cite{Suraud87} from HF calculations.

In our previous works \cite{Gaidarov2011,Gaidarov2012} the
symmetry energy was studied in a wide range of spherical and
deformed nuclei on the basis, as an example, of the Brueckner EDF
of ANM \cite{Brueckner68,Brueckner69}. In these works the
transition from the properties of nuclear matter to those of
finite nuclei was made using the coherent density fluctuation
model \cite{Ant80,AHP}. In Ref.~\cite{Gaidarov2011} a study of the
correlation between the thickness of the neutron skin in finite
nuclei and the nuclear symmetry energy ($s$) for the isotopic
chains of even-even Ni ($A$=74--84), Sn ($A$=124--152) and Pb
($A$=206--214) nuclei, also the neutron pressure ($p_{0}$) and the
asymmetric compressibility ($\Delta K$) for these nuclei was
performed. The calculations were based on the deformed
self-consistent mean-field HF+BCS method using the CDFM and the
Brueckner EDF. The same approaches were used in
Ref.~\cite{Gaidarov2012} for the calculations of the mentioned
quantities of deformed neutron-rich even-even nuclei, such as Kr
($A$=82--120) and Sm ($A$=140-156) isotopes. The numerical results
for $s$, $p_{0}$, and $\Delta K$ for neutron-rich and
neutron-deficient Mg isotopes with $A$=20--36 were presented in
Ref.~\cite{Gaidarov2014}.

The main aim of this work is, apart from the $\rho$-dependence
investigated in our previous works
\cite{Gaidarov2011,Gaidarov2012,Gaidarov2014}, to study also the
temperature dependence of the symmetry energy in finite nuclei. We
focus on the determination of the symmetry energy coefficient, for
which we have explored the local density approximation
\cite{Agrawal2014,Samaddar2007,Samaddar2008, De2012} with some
modifications. In the present paper the thermal evolution of the
symmetry energy coefficient is investigated for Ni, Sn, and Pb
isotopic chains in the interval $T$=0--4 MeV using different model
temperature-dependent local density distributions for these
nuclei. We restrict ourselves to this temperature range because,
in accordance with several findings (e.g., in
Ref.~\cite{Bandyopadhyay90}), the limiting temperature (above
which the nucleus cannot exist as a bound system) has been
evaluated around 4 MeV for finite nuclei with mass number $A\geq
100$. The temperature-dependent densities of these nuclei are
calculated within a self-consistent Skyrme-HFB method using the
cylindrical transformed deformed harmonic-oscillator basis (HFBTHO
densities) \cite{Stoitsov2013,Stoitsov2005}. The kinetic energy
density is calculated either by the HFBTHO code or by the TF
expression up to $T^{2}$ term \cite{Lee2010}. We have used two
parametrizations of the Skyrme force, namely, SLy4 and SkM*, which
were able to give an appropriate description of bulk properties of
spherical and deformed nuclei in the past. In addition, we present
some results for the $^{208}$Pb nucleus with densities obtained
within the ETF method \cite{Brack84,Brack85} and the rigorous
density functional approach (RDFA) \cite{Stoitsov87}. The effect
of temperature on the rms radii of protons and neutrons and the
formation of neutron skin in hot nuclei is also analyzed and
discussed.

This article is organized as follows. In Sec.~II, we give the
theoretical elements to obtain the symmetry energy coefficient
and briefly describe the temperature-dependent nuclear densities.
In Sec.~III, we present the numerical results for hot nuclei
properties and the temperature dependence of the symmetry energy
of finite nuclei. Section IV contains the conclusions.

\section{Theoretical formalism}
\subsection{Temperature-dependent symmetry energy coefficient with Skyrme energy density functional}

For finite systems, different definitions of the symmetry energy
coefficient and its temperature dependence are considered in the
literature. In the present paper we develop an approach to
calculate the symmetry energy coefficient for a specific nucleus
starting with the LDA expression given in
\cite{Samaddar2007,Agrawal2014}:
\begin{equation}
e_{sym}(A,T)=\frac{1}{I^{2}A}\int \rho(r)
e_{sym}[\rho(r),T]\delta^{2}(r)d^{3}r  .
\label{eq:1}
\end{equation}
In Eq.~(\ref{eq:1}) $I=(N-Z)/A$, $e_{sym}[\rho(r),T]$ is the
symmetry energy coefficient at temperature $T$ of infinite nuclear
matter at the value of the total local density
$\rho(r)=\rho_{n}(r)+\rho_{p}(r)$,
$\delta(r)=[\rho_{n}(r)-\rho_{p}(r)]/\rho(r)$ is the ratio between
the isovector  and the isoscalar parts of $\rho(r)$, with
$\rho_{n}(r)$ and $\rho_{p}(r)$ being the neutron and proton local
densities. The symmetry energy coefficient $e_{sym}(\rho,T)$ can
be evaluated in different ways. Following
Refs.~\cite{Agrawal2014,De2012}, we adopt in this work the
definition
\begin{equation}
e_{sym}(\rho,T)=\frac{e(\rho,\delta,T)-e(\rho,\delta=0,T)}{\delta^{2}},
\label{eq:2}
\end{equation}
where $e(\rho,\delta,T)$ is the energy per nucleon in an
asymmetric infinite matter, while $e(\rho,\delta=0,T)$ is that one
of symmetric nuclear matter. These quantities are expressed by
$e={\cal E}(r,T)/\rho$, where ${\cal E}(r,T)$ is the total energy
density of the system. For the Skyrme energy density functional
that we use in our work it has the form:
\begin{eqnarray}
{\cal E}(r,T)&=&\frac{\hbar^{2}}{2m_{n,k}}\tau_{n}+\frac{\hbar^{2}}{2m_{p,k}}\tau_{p} \nonumber \\
&+& \frac{1}{2}t_{0}\left[\left(1+\frac{1}{2}x_{0}\right )\rho^{2}-\left(x_{0}+\frac{1}{2}\right )(\rho_{n}^{2}+\rho_{p}^{2})\right ] \nonumber \\
&+& \frac{1}{12}t_{3}\rho^{\alpha}\left
[\left(1+\frac{x_{3}}{2}\right ) \rho^{2}-\left
(x_{3}+\frac{1}{2}\right )(\rho_{n}^{2}+\rho_{p}^{2})\right ]
\nonumber \\ &+&\frac{1}{16}\left[ 3t_{1}\left
(1+\frac{1}{2}x_{1}\right )-t_{2}\left (1+\frac{1}{2}x_{2}\right
)\right ](\nabla \rho
)^{2} \nonumber \\
&-&\frac{1}{16}\left [ 3t_{1}\left (x_{1}+\frac{1}{2}\right
)+t_{2}\left (x_{2}+\frac{1}{2}\right )\right ] \nonumber \\
&\times & [(\nabla \rho_{n})^{2}+(\nabla \rho_{p})^{2}]+{\cal
E}_{c}(r),
\label{eq:3}
\end{eqnarray}
where for infinite homogeneous nuclear matter only the first three
lines of Eq.~(\ref{eq:3}) contribute. The derivative terms vanish
and the Coulomb term ${\cal E}_{c}$ is neglected. In
Eq.~(\ref{eq:3}) $t_{0}$, $t_{1}$, $t_{2}$, $t_{3}$, $x_{0}$,
$x_{1}$, $x_{2}$, $x_{3}$, and $\alpha$ are the Skyrme parameters.
We use in this work the interactions SkM* \cite{Bartel82} and SLy4
\cite{Chabanat98}. The nucleon effective mass $m_{q,k}$ is defined
through
\begin{eqnarray}
\frac{m}{m_{q,k}(r)}=1+\frac{m}{2\hbar^{2}}\left \{\left [
t_{1}\left (1+\frac{x_{1}}{2}\right )+t_{2}\left
(1+\frac{x_{2}}{2}\right )\right ]\rho \right. \nonumber \\
+\left. \left[ t_{2}\left (x_{2}+\frac{1}{2}\right )-t_{1}\left
(x_{1}+\frac{1}{2}\right )\right ]\rho_{q} \right \},
\label{eq:4}
\end{eqnarray}
with $q=(n,p)$ referring to neutrons or protons. The dependence on
temperature of ${\cal E}(r,T)$ [Eq.~(\ref{eq:3})] and
$m/m_{q,k}(r)$ [Eq.~(\ref{eq:4})] comes from the $T$-dependence of
the densities and kinetic energy densities.

A self-consistent approach based on the simultaneous treatment of
temperature-dependent density distributions and kinetic energy
density is related to the finite temperature formalism for the HFB
method. In it the nuclear Skyrme-HFB problem is solved by using
the transformed harmonic-oscillator basis \cite{Stoitsov2013}. The
HFBTHO code based on the mentioned approach is used in our
numerical calculations.

The HFBTHO code solves the finite temperature HFB equations
assuming axial and time-reversal symmetry. These equations are
formally equivalent to the HFB equations at $T=0$ if the
expressions of the density matrix $\rho$ and pairing tensor
$\kappa$ are redefined as
\begin{eqnarray}
\rho &=& U f U^\dagger + V^\ast (1-f) V^T \, , \nonumber \\
\kappa &=& U f V^\dagger + V^\ast (1-f)U^T \, ,
\label{rhokappa}
\end{eqnarray}
%
%
%
where $U$ and $V$ are the matrices of the Bogoliubov
transformation (here $T$ means transpose) and $f$ is the
temperature-dependent Fermi-Dirac factor given by
\begin{equation}
f_i = \left( 1+ e^{{E_i}/k_BT} \right) ^{-1} \, .
\end{equation}
In this expression $E_i$ is the quasiparticle energy of the state
$i$ and $k_B$ is the Boltzmann constant. In HFBTHO the Fermi level
$\lambda$ is determined at each iteration from the conservation of
particle number in BCS approach \cite{Stoitsov2013},
\begin{equation}
N(\lambda ) = \sum_i  \left[ v_i(\lambda)^2 + f_i (\lambda) \left(  u_i (\lambda)^2 - v_i(\lambda) ^2 \right) \right] \, , \label{nlambda}
\end{equation}
where the BCS occupations are given by
\begin{equation}
v_i^2 = \frac{1}{2} \left[ 1- \frac{e_i -\lambda}{E_i^{BCS}}  \right] \, , \quad
 u_i ^2 = 1-  v_i ^2 \, ,
\end{equation}
and $E_i^{BCS}= \left[ (e_i-\lambda)^2+\Delta_i ^2  \right] ^{1/2}$. Note that at $T=0$ the Fermi-Dirac factors are zero and one recovers the usual expressions for
$\rho$ and $\kappa$ in Eq. (\ref{rhokappa}) and for the number of particles in Eq. (\ref{nlambda}).

\subsection{Temperature-dependent kinetic energy density}

There exist various methods to obtain the kinetic energy density
$\tau_{q}(r,T)$ entering the expression for ${\cal E}(r,T)$
[Eq.~(\ref{eq:3})]. One of them is, as mentioned above, to use the
HFBTHO code. Another way is to use the TF approximation
adopted in Ref.~\cite{Agrawal2014}, or an extension of the TF expression
up to $T^{2}$ terms \cite{Lee2010}:
\begin{eqnarray}
\tau_{q}(r,T)&=&\frac{2m}{\hbar^{2}}\varepsilon_{K_{q}}=\frac{3}{5}(3\pi^{2})^{2/3}
\nonumber \\
&\times & \left [
\rho_{q}^{5/3}+\frac{5\pi^{2}m_{q}^{2}}{3\hbar^{4}}\frac{1}{(3\pi^{2})^{4/3}}\rho_{q}^{1/3}T^{2}\right
].
\label{eq:5}
\end{eqnarray}
In Eq.~(\ref{eq:5}) the first term in square brackets is the
degenerate limit at zero temperature and the $T^{2}$ term is the
finite-temperature correction. By using the approximate expression
(\ref{eq:5}) for the kinetic energy density, Lee and Mekjian
performed calculations of the volume and surface symmetry energy
coefficients for finite nuclei in Ref.~\cite{Lee2010} showing that
the surface symmetry energy term is the most sensitive to the
temperature while the bulk energy term is the least sensitive. In
the present work we calculate the kinetic energy density using the
self-consistent Skyrme-HFB method and the HFBTHO code. Also, for a
comparison we present the results when using $\tau_{q}(r,T)$ from
Eq.~(\ref{eq:5}).

\subsection{Temperature-dependent densities}

In our work the local density distributions are calculated by the
HFBTHO code \cite{Stoitsov2013}. The $T$-dependent proton and
neutron densities $\rho_{q}({\vec r},T)$ normalized by
\begin{equation}
\int \rho_{q}({\vec r},T) d{\vec r} = Q,   \;\;\;\;\;\;  Q=Z,N
\label{eq:5a}
\end{equation}
determine the corresponding mean square radii
\begin{equation}
<R_{\rm q}^2> =\frac{ \int r^2\rho_{\rm q}({\vec r},T)d{\vec
r}} {\int \rho_{\rm q}({\vec r},T)d{\vec r}} \, ,
\label{eq:5b}
\end{equation}
the rms radii
\begin{equation}
R_{\rm q}=<R_{\rm q}^2> ^{1/2} \, ,
\label{eq:5c}
\end{equation}
and the neutron skin thickness which is usually characterized by
the difference of the neutron and proton rms radii:
\begin{equation}
\Delta R=R_{\rm n}-R_{\rm p}.
\label{eq:5d}
\end{equation}

In addition, two other density distributions of $^{208}$Pb
\cite{Antonov89}, namely the Fermi-type density determined within
the ETF method \cite{Brack84,Brack85} and the symmetrized-Fermi
local density obtained within the rigorous density functional
approach (RDFA) \cite{Stoitsov87}, are used. The density within
the ETF method \cite{Brack84,Brack85} which is the semi-classical
limit of the temperature-dependent Hartree-Fock (THF) theory
\cite{Brack74} has the form:
\begin{equation}
\rho_{ETF}(r,T)=\rho_{0}(T)\left \{1+\exp \left [
\frac{r-R(T)}{\alpha(T)}\right ]\right \}^{-\gamma(T)}.
\label{eq:6}
\end{equation}
The temperature-dependent local density parameters $\rho_{0}$,
$R$, $\alpha $ and $\gamma $ are obtained for the nucleus
$^{208}$Pb with the SkM* effective force. The local densities
(\ref{eq:6}) reproduce the averaged THF results up to temperature
$T$=4 MeV \cite{Brack74}. The symmetrized-Fermi local density
distribution determined for the same nucleus within the RDFA
\cite{Stoitsov87} is
\begin{equation}
\rho_{SF}(r,T)=\rho_{0}(T)\frac{\sinh[R(T)/b(T)]}{\cosh[R(T)/b(T)]+\cosh[r/b(T)]}.
\label{eq:7}
\end{equation}
The temperature-dependent local density parameters $\rho_{0}$,
$R$, and $b$ are obtained with the SkM effective force up to
$T$=10 MeV. As it has been demonstrated in \cite{Stoitsov87}, the
RDFA reproduces almost exactly the THF results \cite{Bonche} up to
temperatures $T$=8 MeV above which the nucleus is unstable with
respect to the THF calculations \cite{Bonche}.

\subsection{Relationships for calculations of $T$-dependent symmetry energy
coefficient}

As mentioned in Sec.~II~A, in the present work we use the approach
given by Eqs. (\ref{eq:1}) and (\ref{eq:2}), as well as the
$T$-dependent Skyrme EDF [Eq.~(\ref{eq:3})] to calculate the
symmetry energy coefficient. Here we note the specific problem
that arises, namely how to calculate the term $e(\rho,\delta=0,T)$
of Eq.~(\ref{eq:2}) that is responsible for the contribution of
the energy per particle of symmetric nuclear matter. One of the
expressions shown in Ref.~\cite{De2012} to calculate $e_{sym}(T)$
for a nucleus with mass number $A$ is in the spirit of the
liquid-drop model and has the form:
\begin{equation}
e_{sym}(T) =[e(N,Z,T)-e(A/2,A/2,T )]/X^{2},
\label{eq:5e}
\end{equation}
where $X=(N - Z)/A$ is the asymmetry parameter. Eq.~(\ref{eq:5e})
is valid when the energy per particle of the nucleus $e$ does not
contain the Coulomb contribution. As pointed out in
Ref.~\cite{De2012} in the cases of relatively heavy nuclei, the
stable systems are usually isospin-asymmetric and then, the
definition given by Eq.~(\ref{eq:5e}) may not be operative. The
suggested expression in \cite{Agrawal2014,De2012} is
\begin{equation}
e_{sym}(T) =[e(A,X_{1},T)-e(A,X_{2},T )]/(X_{1}^{2}-X_{2}^{2}),
\label{eq:5f}
\end{equation}
where $X_{1}$ and $X_{2}$ are the asymmetry parameters of the
nuclear pair. As it has been concluded in \cite{De2012} the value
of $e_{sym}(T)$ from Eqs.~(\ref{eq:5e}) and (\ref{eq:5f}) depends
on the choice of the nuclear pair and, thus, its value is not
unambiguous for a particular nucleus.

Therefore, in our study aiming to investigate the temperature
dependence of $e_{sym}$ within a given isotopic chain, we
introduce other definitions of $e_{sym}(A,T)$ in LDA that, in our
opinion, would be more appropriate in this case. They concern
namely the above mentioned problem of calculating the term
$e(\rho,\delta=0,T)$ of Eq.~(\ref{eq:2}) for symmetric nuclear
matter. In our LDA approach the latter is simulated by considering
the $N=Z=A/2$ nucleus, but we analyze two possibilities. First, on the basis of
Eqs.~(\ref{eq:1}) and (\ref{eq:2}) with $e={\cal E}(r)/\rho$, we
present the integrand of the right-hand side of the following
expression for $I^{2} e_{sym}(A,T)$ as a difference of two terms
with transparent physical meaning:
\begin{widetext}
\begin{equation}
I^{2} e_{sym}(A,T)=\int d{\vec r}\left [ \frac{{\cal
E}(\rho_{A}(r),\delta,T)}{A}-\frac{{\cal
E}(\rho_{A1}(r),\delta=0,T)}{A1} \right ],
\label{eq:8}
\end{equation}
\end{widetext}
in which the first one corresponds to the energy per volume and particle of nuclear matter ${\cal E}(\rho_{A}(r),\delta,T)/A$ with a density $\rho_{A}(r)$ equal to
that of the considered nucleus with $A$ nucleons, $Z$ protons and $N$ neutrons from the given isotopic chain. The second term ${\cal E}(\rho_{A1}(r),\delta=0,T)/A1$
is the analogous for the isotope with $A1=2Z$ ($N1=Z=A1/2$). For example, for the Ni isotopic chain the nucleus $A1$ is the double-closed shell nucleus $^{56}$Ni
($Z=N1=28$), while for the Sn isotopic chain the nucleus $A1$ is the double-closed shell nucleus $^{100}$Sn ($Z=N1=50$) and both $^{56}$Ni and $^{100}$Sn isotopes
play a role of reference nuclei.

Our second new definition of $e_{sym}(A,T)$ is based on the
expression (\ref{eq:5e}) that is for finite nuclei. The latter
allows us, using the LDA, to present $e_{sym}(A,T)$ in the form:
\begin{widetext}
\begin{equation}
I^{2} e_{sym}(A,T)=\int \frac{d{\vec r}}{A} \left [ {\cal
E}(\rho_{A}(r),\delta,T)-{\cal
E}(\rho_{\bar{A}}(r),N=\bar{A}/2,Z=\bar{A}/2,\delta=0,T)
\right ],
\label{eq:8a}
\end{equation}
\end{widetext}
in which the mass number $\bar{A}=A$ is the same, but with different nucleon
content, $A(Z,N)$ and $\bar{A}(Z=\bar{A}/2,N=\bar{A}/2)$.
This consideration requires the even-even nucleus with
$N=Z=\bar{A}/2$ to be bound.

In the calculations (with results presented in Sec.~III~B) the
$T$-dependent densities and kinetic energy densities are
calculated by using Eqs.~(\ref{eq:1})--(\ref{eq:4}), as well as
the HFBTHO code. For a comparison, the results obtained by using
TF expression with $T^{2}$ term for the $T$-dependent kinetic
energy densities [Eq.~(\ref{eq:5})], as well as those obtained by
using $T$-dependent densities from the ETF method
[Eq.~(\ref{eq:6})] and the RDFA [Eq.~(\ref{eq:7})] for $^{208}$Pb
nucleus are also presented in Sec.~III~B.

\section{Results for Ni, Sn, and Pb isotopic chains and discussion}
\subsection{Temperature-dependent densities, nuclear radii, and
neutron skins}

\begin{figure*}
\includegraphics[width=0.48\linewidth]{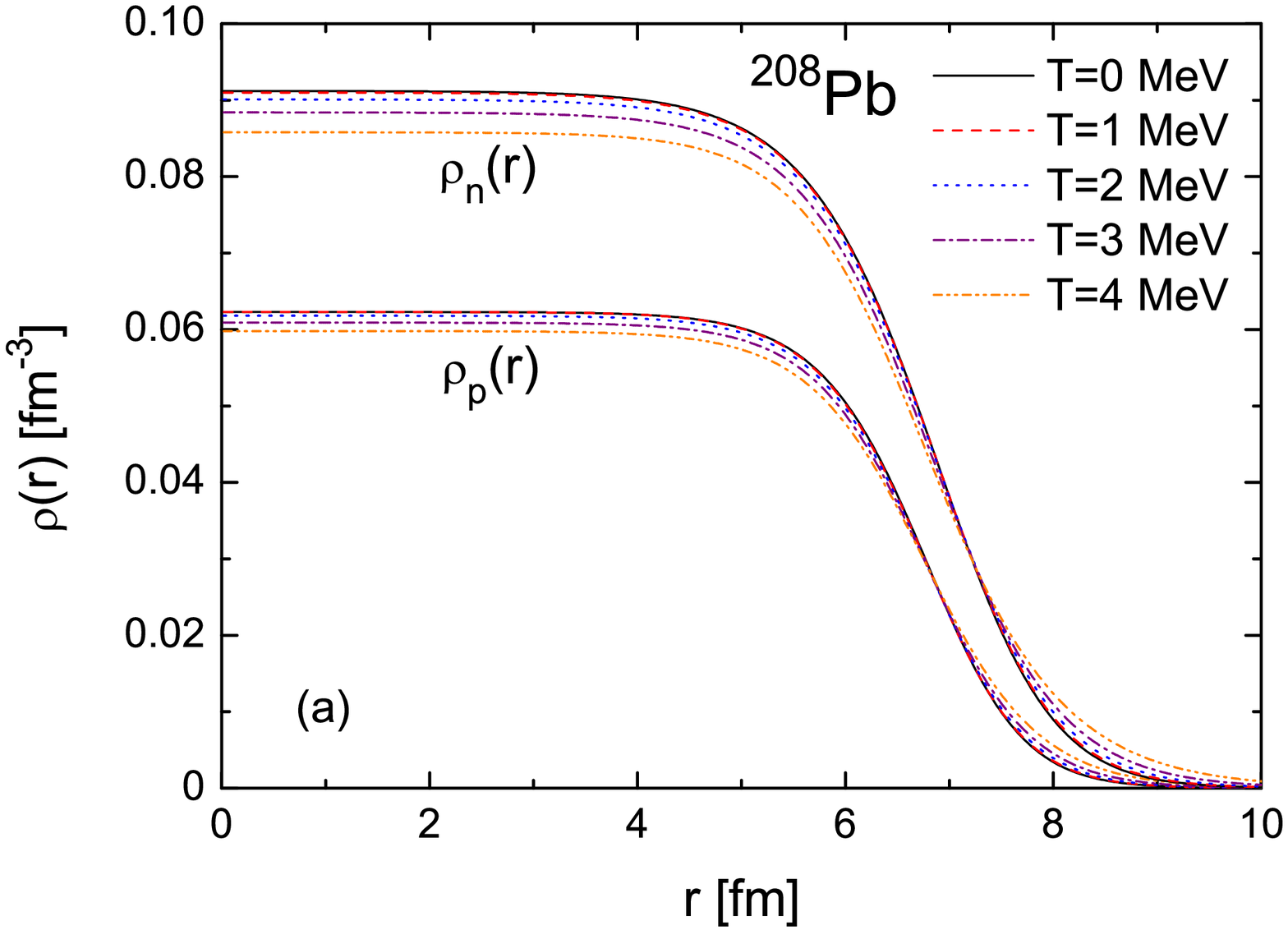}
\includegraphics[width=0.48\linewidth]{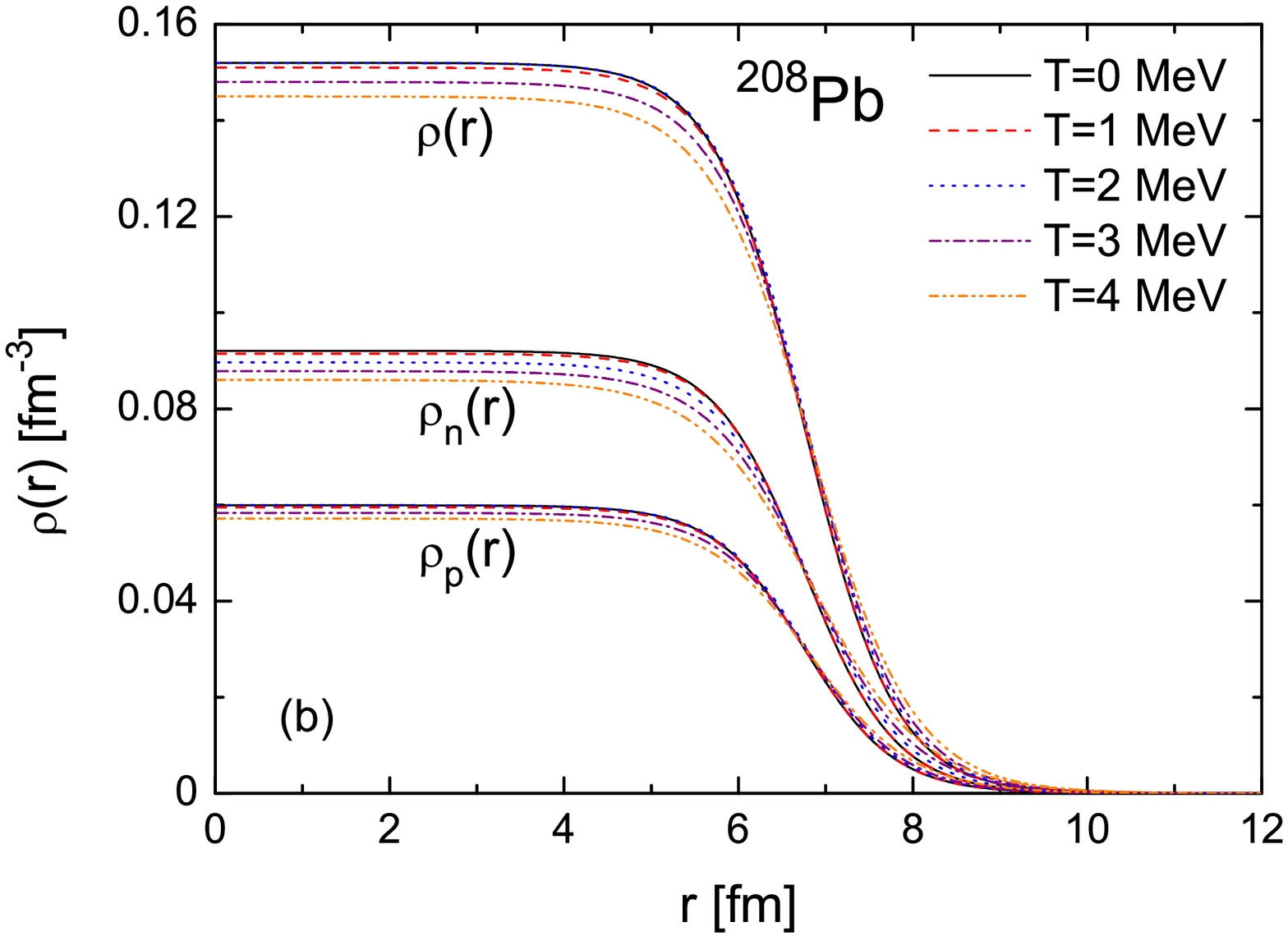}
\caption{(Color online) Proton $\rho_{p}(r)$ and neutron
$\rho_{n}(r)$ local density distributions of $^{208}$Pb obtained
within the ETF method (a) and RDFA (b) for temperatures $T$=0 MeV
(black solid line), $T$=1 MeV (red dashed line), $T$=2 MeV (blue
dotted line), $T$=3 MeV (purple dash-dotted line), and $T$=4 MeV
(yellow dash-double-dotted line). The total density distribution
$\rho(r)$ of $^{208}$Pb obtained within the RDFA is also
presented.
\label{fig1}}
\end{figure*}

\begin{figure*}
\includegraphics[width=0.48\linewidth]{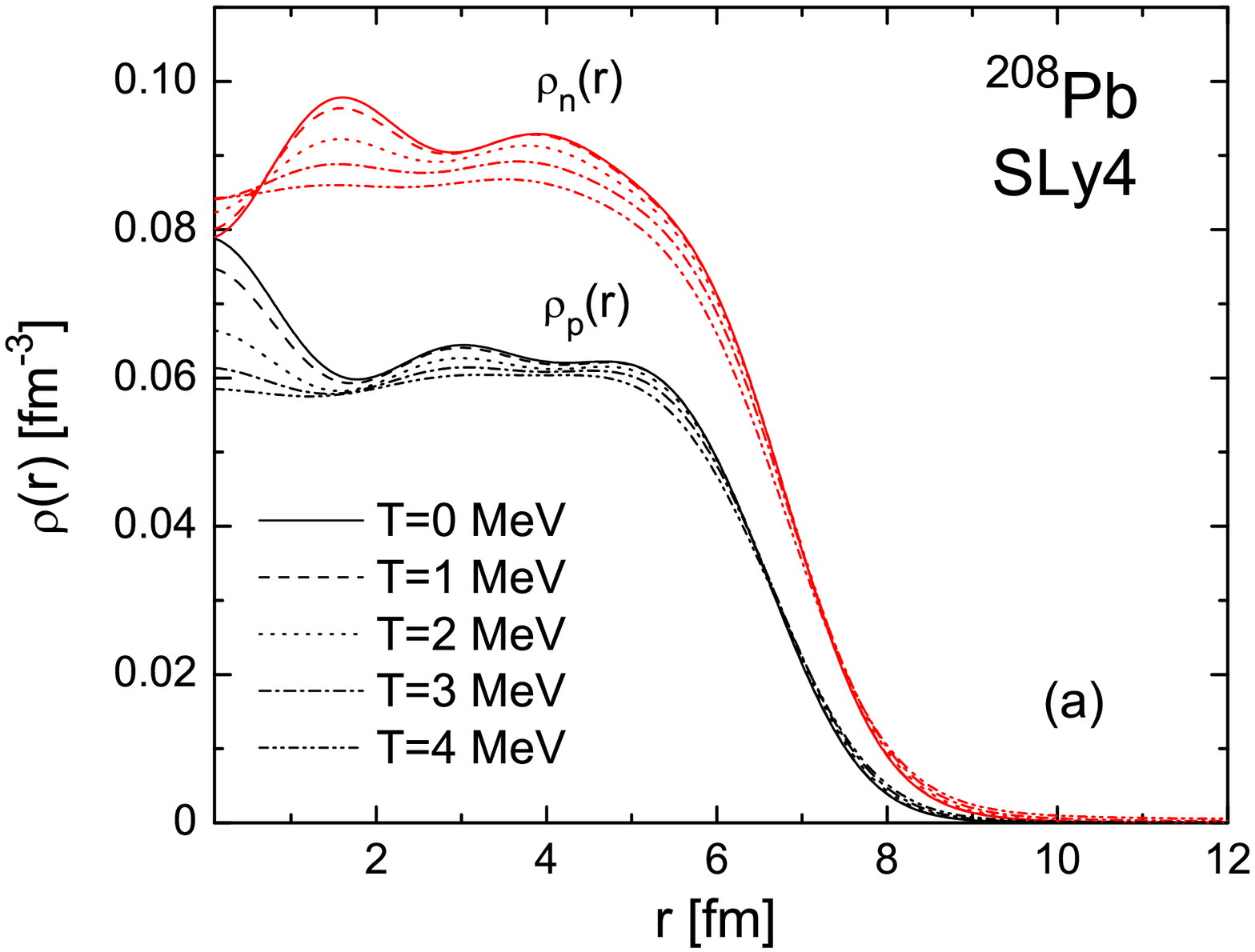}
\includegraphics[width=0.48\linewidth]{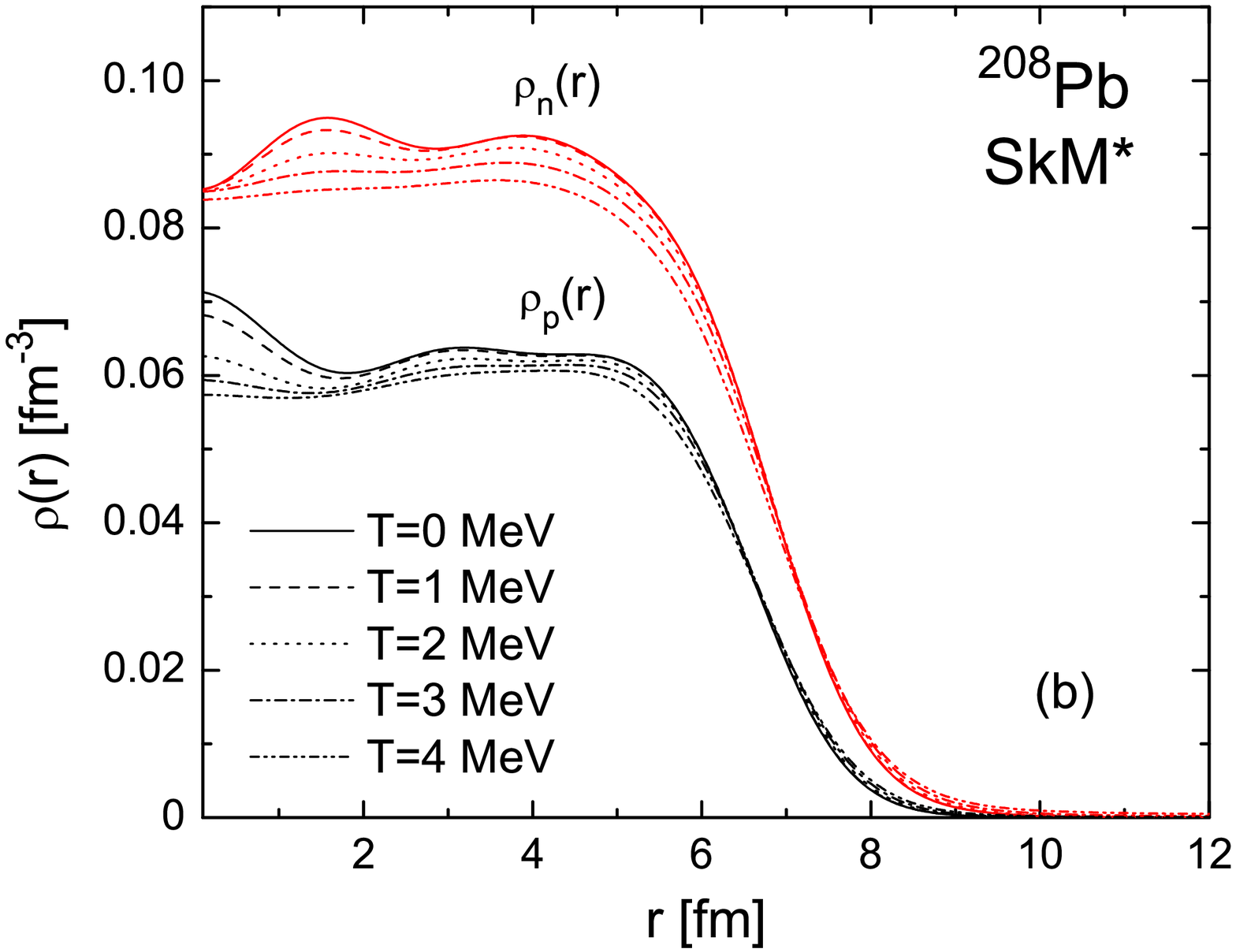}
\caption{(Color online) Proton and neutron local density
distributions of $^{208}$Pb obtained within the HFBTHO method
\protect\cite{Stoitsov2013} with SLy4 (a) and SkM* (b) forces.
Five different curves for protons (in black) and neutrons (in red)
represent the results for the corresponding densities for
temperatures $T$=0 (solid line), $T$=1 MeV (dashed line), $T$=2
MeV (dotted line), $T$=3 MeV (dash-dotted line), and $T$=4 MeV
(dash-double-dotted line).
\label{fig2}}
\end{figure*}

We start our analysis by studying the local density distributions
$\rho(r)$ and their changes with respect to the temperature. The
results for these densities of the nucleus $^{208}$Pb obtained
within the ETF method [Eq.~(\ref{eq:6})] and RDFA
[Eq.~(\ref{eq:7})], as well as the Skyrme HFB method, are given in
Figs.~\ref{fig1} and \ref{fig2}, respectively. In addition to the
proton and neutron densities, normalized to $Z$=82 and $N$=126,
respectively, that are presented in the left panel of
Fig.~\ref{fig1}, we give also in the right panel of the same
figure the total local density of $^{208}$Pb normalized to
$A$=208. It can be seen that ETF method and RDFA yield densities
that have smooth behavior with $r$ at any temperature $T$ although
the RDFA, in contrast to ETF method, incorporates the THF shell
effects \cite{Antonov89}. Figure~\ref{fig1} also shows that with
increasing temperature all type of densities decrease in the
central part of the nucleus. This decrease is stronger for the
neutron distributions of $^{208}$Pb. The proton and neutron local
density distributions of $^{208}$Pb obtained within the Skyrme HFB
method in Fig.~\ref{fig2} have somewhat different behavior. The
same trend with the increase of the temperature can be observed,
but in this case the HFBTHO densities exhibit a stronger
$T$-dependence. At the same time, it is observed that the nuclear
surface becomes more diffuse with increasing $T$, while a similar
reduction of the densities at the center of the nucleus shown in
Fig.~\ref{fig1} takes place. This is a natural consequence of the
weakness of the shell effects with increasing $T$.


\begin{figure*}
\includegraphics[width=0.48\linewidth]{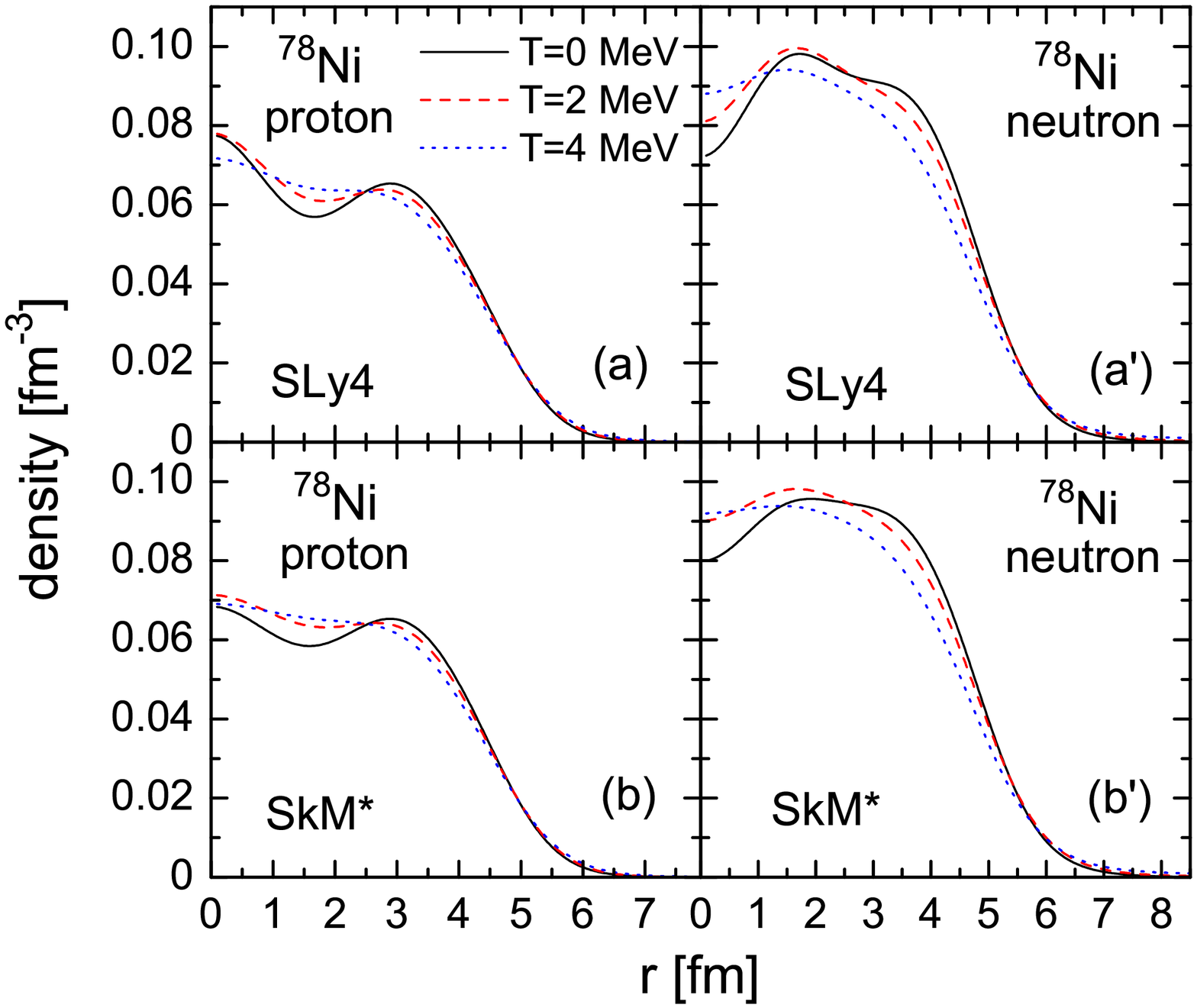}
\includegraphics[width=0.48\linewidth]{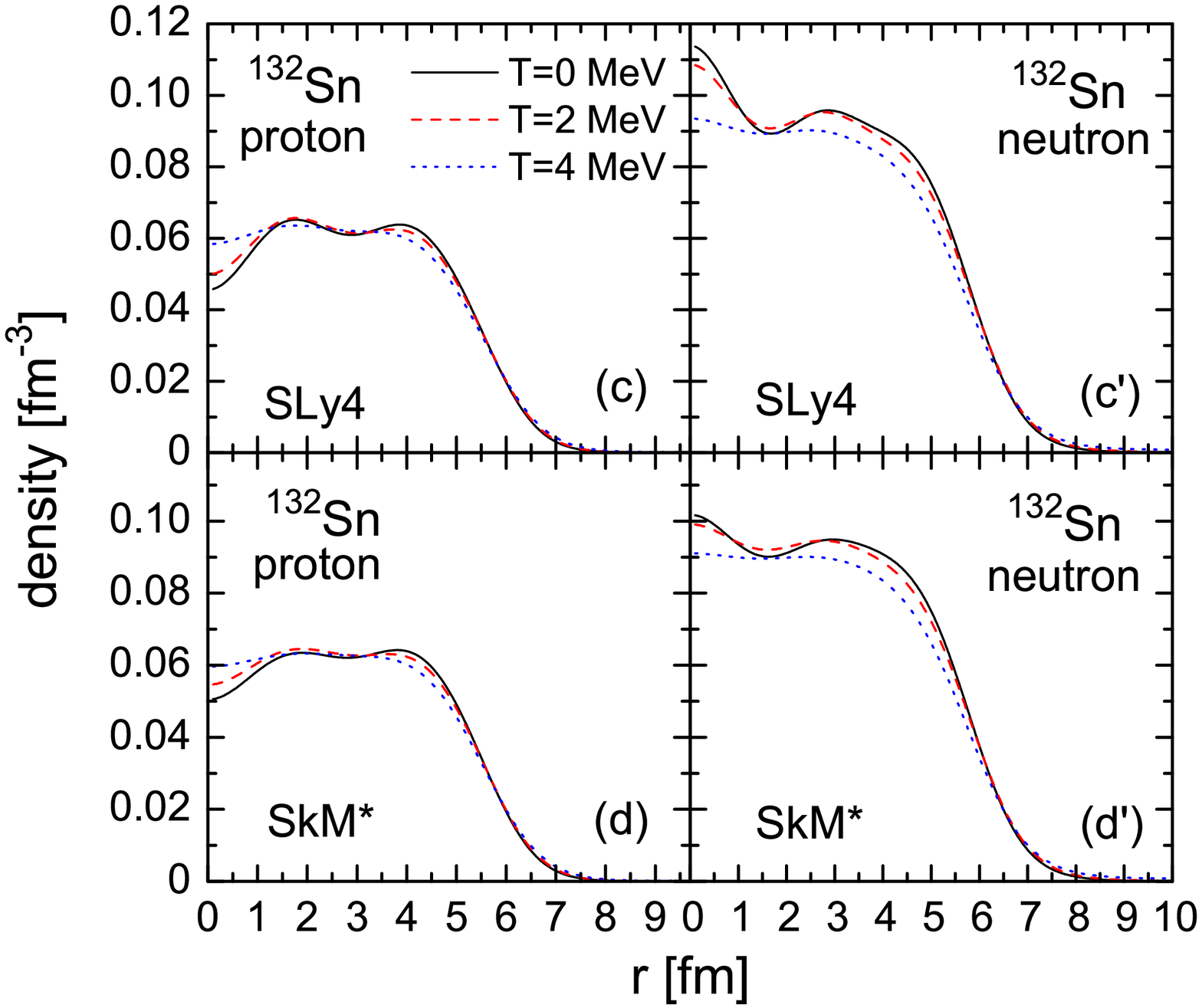}
\caption{(Color online) HFBTHO density distributions of protons
and neutrons for $^{78}$Ni [(a), (a$^{\prime}$), (b),
(b$^{\prime}$)] and $^{132}$Sn [(c), (c$^{\prime}$), (d),
(d$^{\prime}$)] at $T$=0 MeV (solid line), $T$=2 MeV (dashed
line), and $T$=4 MeV (dotted line) obtained using the SLy4 and
SkM* parametrizations.
\label{fig3}}
\end{figure*}

In Fig.~\ref{fig3} we display as examples the density
distributions of protons and neutrons for double-magic $^{78}$Ni
and $^{132}$Sn nuclei at $T$=0, 2, and 4 MeV obtained by using the
SLy4 and SkM* parametrizations within the HFB method. It can be
seen from Fig.~\ref{fig3} that the differences between the curves
corresponding to the three temperatures are smaller in these
nuclei in comparison with the case of $^{208}$Pb shown in
Fig.~\ref{fig2}. The tendency in the behavior of proton and
neutron densities of $^{78}$Ni and $^{132}$Sn obtained with a
given Skyrme force (SLy4 or SkM*) is similar. For example, the use
of both parametrizations leads to a depression of the proton
densities in the interior of $^{132}$Sn [Figs.~\ref{fig3}(c) and
\ref{fig3}(d)] being larger at zero temperature and to a growth in
the same region for the neutron densities, but in an opposite
direction relative to $T$ [Figs.~\ref{fig3}(c$^{\prime}$) and
\ref{fig3}(d$^{\prime}$)]. As a consequence, a spatial extension
of both densities at the surface region is observed with the
increase of $T$. Namely this region is responsible for the
emergence of a neutron skin (e.g., Ref.~\cite{Sarriguren2007}).

\begin{figure}
\centering
\includegraphics[width=85mm]{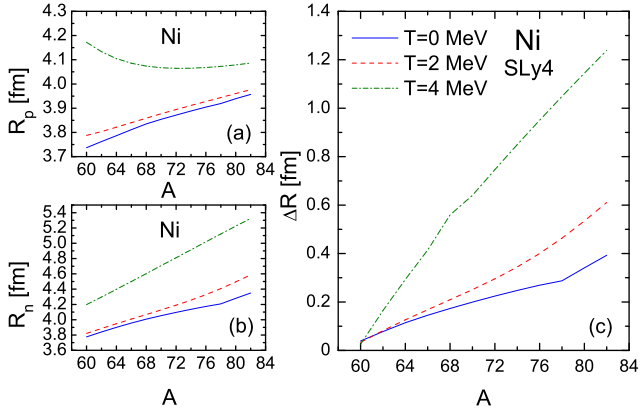}
\caption[]{(Color online) Mass dependence of the proton $R_{p}$
(a) and neutron $R_{n}$ (b) radius of the Ni isotopes ($A$=60--82)
calculated with SLy4 interaction at $T=0$ MeV (solid line), $T=2$
MeV (dashed line), and $T=4$ MeV (dash-dotted line). Neutron skin
thickness $\Delta R$ as a function of $A$ (c) for the Ni isotopes.
\label{fig4}}
\end{figure}

\begin{figure}
\centering
\includegraphics[width=85mm]{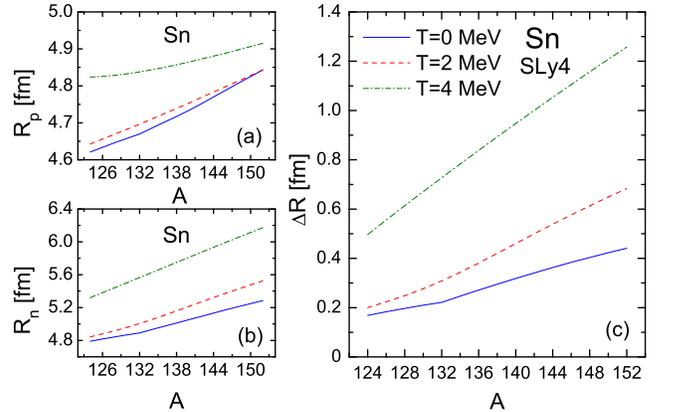}
\caption[]{(Color online) Same as in Fig.~\ref{fig4}, but for Sn
isotopes ($A$=124--152).
\label{fig5}}
\end{figure}

\begin{figure}
\centering
\includegraphics[width=85mm]{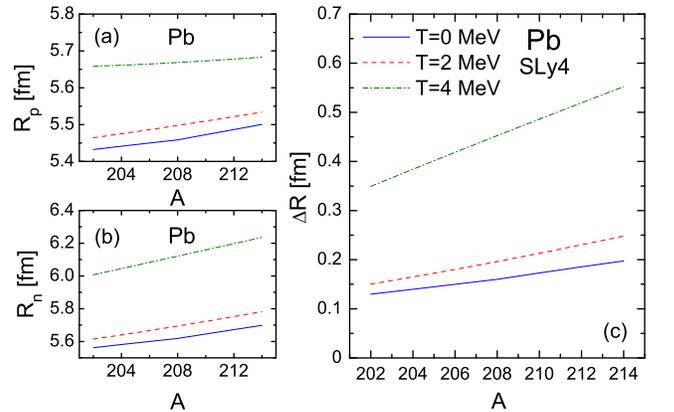}
\caption[]{(Color online) Same as in Fig.~\ref{fig4}, but for Pb
isotopes ($A$=200--214).
\label{fig6}}
\end{figure}

We show in Figs.~\ref{fig4}--\ref{fig6} the neutron and proton rms
radii [Eq.~(\ref{eq:5c})] and their difference known as the
neutron-skin thickness [Eq.~(\ref{eq:5d})] as a function of the
mass number $A$ for Ni ($A$=60--82), Sn ($A$=124--152), and Pb
($A$=202--214) isotopic chains, respectively, calculated by using
SLy4 force. First, it can be seen that the proton rms radii for
all cases increase more slowly than the neutron ones, which is
valid for all the isotopic chains and temperatures. This is
naturally expected in isotopic chains where the number of protons
remains fixed. In addition, while the results of both radii at
$T=0$ and $T=2$ MeV are close to each other with increasing $A$,
one can see a steep increase of their values when the nucleus
become very hot ($T$=4 MeV). In the case of Pb isotopes there is
almost no change of the proton radius within the chain at $T=4$
MeV [Fig.~\ref{fig6}(a)]. The neutron rms radii for the same chain
tend to increase [see Fig.~\ref{fig6}(b)], but not so rapidly as
they increase for the Ni and Sn isotopes (Figs.~\ref{fig4} and
\ref{fig5}). As can be seen from Figs.~\ref{fig4}(c),
\ref{fig5}(c), and \ref{fig6}(c), the neutron-skin thickness
exhibits the same trend as the rms radii. It grows significantly
with the increase of $T$ being much larger at $T$=4 MeV than at
lower temperatures $T$=0, 2 MeV.

The mechanism of formation of neutron skin in tin isotopes has been
studied in Ref.~\cite{Yuksel2014}, where the
changes in the neutron skin was attributed mainly to the effect of
temperature on the occupation probabilities of the single-particle
states around the Fermi level. In \cite{Yuksel2014} a more limited
Sn isotopic chain up to $^{120}$Sn was considered. Our results for
larger $A$ in this chain (from $A=124$ to $A=152$) also show a
slow increase of the neutron skin size. The enhancement of the proton and
neutron radii at high temperatures leads to a rapid increase of
the neutron skin size. We would like to note that at zero
temperature, the use of HFBTHO temperature-dependent densities in
the present approach confirms the observation in our previous work
\cite{Sarriguren2007} (where the densities were calculated within
a deformed Skyrme HF+BCS approach), namely that a pronounced neutron skin can
be expected at $A>132$ in Sn and $A>74$ in Ni isotopes.


\begin{figure}
\centering
\includegraphics[width=78mm]{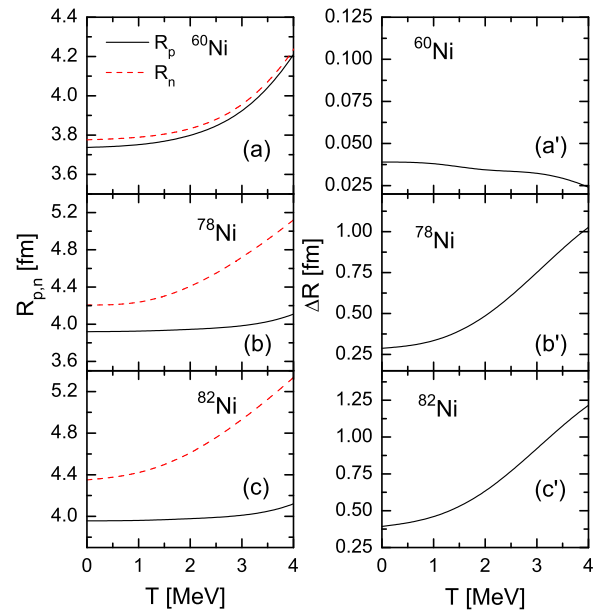}
\caption[]{(Color online) Left: Proton $R_{p}$ (solid line) and
neutron $R_{n}$ (dashed line) radius of $^{60}$Ni, $^{78}$Ni, and
$^{82}$Ni isotopes with respect to the temperature $T$ calculated
with SLy4 interaction. Right: Neutron skin thickness $\Delta R$
for the same Ni isotopes as a function of $T$.
\label{fig7}}
\end{figure}

\begin{figure}
\centering
\includegraphics[width=78mm]{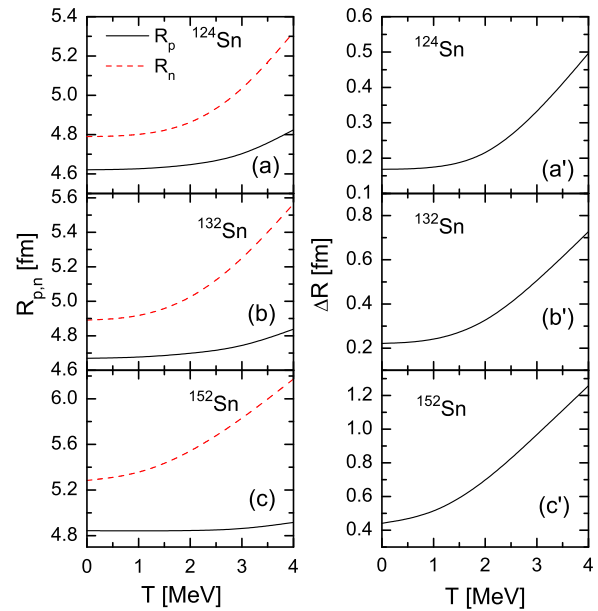}
\caption[]{(Color online) Same as in Fig.~\ref{fig7}, but for
 $^{124}$Sn, $^{132}$Sn, and $^{152}$Sn isotopes.
\label{fig8}}
\end{figure}

\begin{figure}
\centering
\includegraphics[width=78mm]{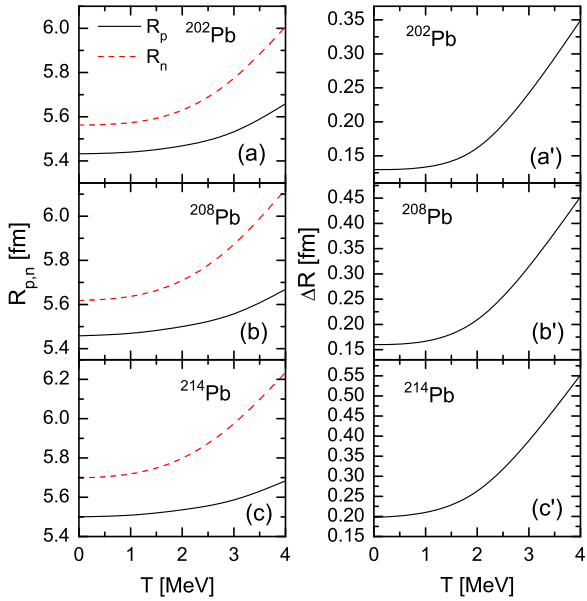}
\caption[]{(Color online) Same as in Fig.~\ref{fig7}, but for
 $^{202}$Pb, $^{208}$Pb, and $^{214}$Pb isotopes.
\label{fig9}}
\end{figure}

The results for the proton and neutron radii and their difference
(neutron-skin thickness) as a function of the temperature $T$ for
selected $^{60}$Ni, $^{78}$Ni, and $^{82}$Ni isotopes, are
illustrated in Fig.~\ref{fig7}. The calculations are made by using
SLy4 parametrization. In addition, similar plots with results for
three tin ($^{124}$Sn, $^{132}$Sn, $^{152}$Sn) and lead
($^{202}$Pb, $^{208}$Pb, $^{214}$Pb) isotopes are presented in
Figs.~\ref{fig8} and \ref{fig9}, respectively. In the temperature
range $T$=0--4 MeV considered in the present work, we find a very
slow increase of the proton radius compared to the rapid increase
of the neutron radius with the temperature. As it is seen from
Fig.~\ref{fig7}(a), only for $^{60}$Ni nucleus both proton and
neutron rms radii are very similar and behave similarly with
temperature. In fact, the dependence of the neutron skin thickness
with temperature in $^{60}$Ni is very small and we observe only a
tiny effect compatible with an almost null skin thickness [we note
the small scale in $^{60}$Ni in Fig.~\ref{fig7}(a$^{\prime}$) in
comparison with those of $^{78}$Ni and $^{82}$Ni in
Figs.~\ref{fig7}(b$^{\prime}$) and (c$^{\prime}$), respectively].
Here we would like to note that the use of SkM* interaction leads
to results for the proton and neutron radii, as well as for the
neutron skin thickness of the considered isotopes, very similar to
those obtained by using of SLy4 Skyrme force and presented in
Figs.~\ref{fig4}--\ref{fig9}.

The temperature dependence of rms radii obtained in this work for
$^{208}$Pb can be compared to that shown in Ref.~\cite{Zhang2014}.
In the latter work properties of hot nuclei have been studied
within the relativistic TF approximation and different RMF
parametrizations were tested. The temperature dependence of the
proton radius agrees well with that of Ref.~\cite{Zhang2014},
which in the range below $T=4$ MeV is quite independent of the RMF
parametrization used. On the other hand, the temperature
dependence of the neutron radius in \cite{Zhang2014} is more
sensitive to the RMF parametrization used. The dependence on $T$
of the neutron radii in our calculations is more pronounced,
increasing with $T$ much faster than those in
Ref.~\cite{Zhang2014}. As a result, the neutron skin thickness,
which is rather flat in \cite{Zhang2014}, increases more rapidly
with $T$ in our calculations. It turns out that we get values of
the neutron skin thickness at zero temperature (0.16 fm) similar
to those obtained in \cite{Zhang2014} by using FSU
parametrization. As it has been pointed out in
Ref.~\cite{Sarriguren2007}, the RMF results for $\Delta R$
systematically overestimate the Skyrme HF results. This is
confirmed by the larger values of the neutron-skin thickness of
$^{208}$Pb obtained in Ref.~\cite{Zhang2014} when using NL3 and
TM1 models.

%

\subsection{Temperature dependence of the symmetry energy coefficient}

\begin{figure}
\centering
\includegraphics[width=80mm]{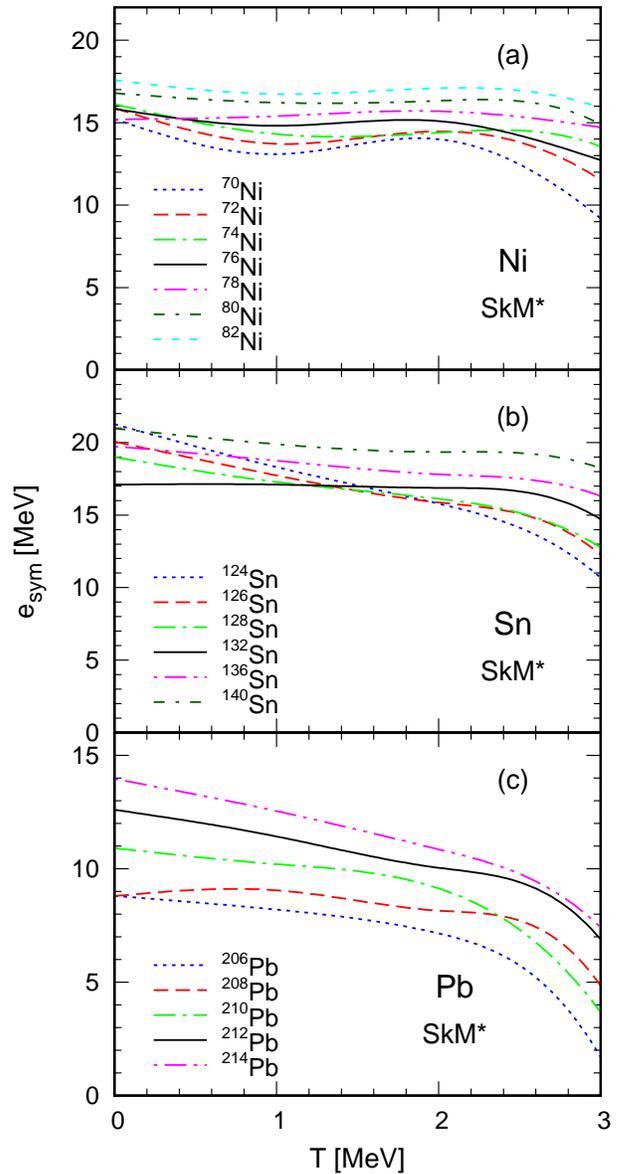}
\caption[]{(Color online) Temperature dependence of the symmetry
energy coefficient $e_{sym}$ obtained by using Eq.~(\ref{eq:8})
for several nuclei from Ni ($A$=70--82) (a), Sn ($A$=124--140)
(b), and Pb ($A$=206--214) (c) isotopic chains with SkM* force.
The nucleon densities and kinetic energy densities used to
calculate $e_{sym}$ are consistently derived from HFBTHO code.
\label{fig10}}
\end{figure}

In understanding the symmetry energy coefficient $e_{sym}$ for finite nuclear systems and their thermal evolution, some ambiguities about their proper definition
could be noted. First, we use the new definition of the symmetry energy coefficient $e_{sym}$ given by Eq.~(\ref{eq:8}) and the results for several nuclei from the
three isotopic chains calculated with SkM* interaction are presented in Fig.~\ref{fig10}. They are obtained by simultaneous consistent treatment of both
$T$-dependent nucleon densities and kinetic energy densities within the HFB method and computed by the HFBTHO code. As noted in subsection~II~D, there exist
difficulties in the calculations of the term $e(\rho,\delta=0,T)$ of Eq.~(\ref{eq:2}) for symmetric nuclear matter, namely, of using the reference case $\delta=0$
when the nucleus with $Z=N1$ is unbound. Keeping this in mind, as an attempt, for Ni and Sn isotopes we take as reference nuclei ($A1$) the nuclei $^{56}$Ni
($Z=N1=28$) and $^{100}$Sn ($Z=N1=50$), respectively. The case of the Pb isotopic chain is even more difficult because the eventual nucleus of reference with
$Z=N1=82$ is clearly unbound and there do not exist appropriate bound nuclei for the purpose. As a way to overcome this difficulty, we try in this case to use again
the $^{100}$Sn as a reference nucleus with $Z=N1=50$, normalized with $A1=100$ in Eq.~(\ref{eq:8}).

The symmetry energy coefficient exhibits almost flat behavior for the
double-magic $^{78}$Ni and $^{132}$Sn nuclei. Here we would like
also to emphasize that if one extends the temperature range, the
values of $e_{sym}$ may become negative. This fact has been
already discussed in the literature, for instance in
Refs.~\cite{De2012,Zhang2014}, where a subtraction procedure has
been employed for modelling the hot nucleus. The negativity of
$e_{sym}$ at high temperatures violates the general understanding
of the symmetry energy. Generally, however, in our opinion the
expression (\ref{eq:8}) is reliable, particularly when considering
isotopic chains, but obviously the question about the proper
definition of the symmetry energy coefficients for finite nuclei
still remains open.

\begin{figure}[b]
\centering
\includegraphics[width=80mm]{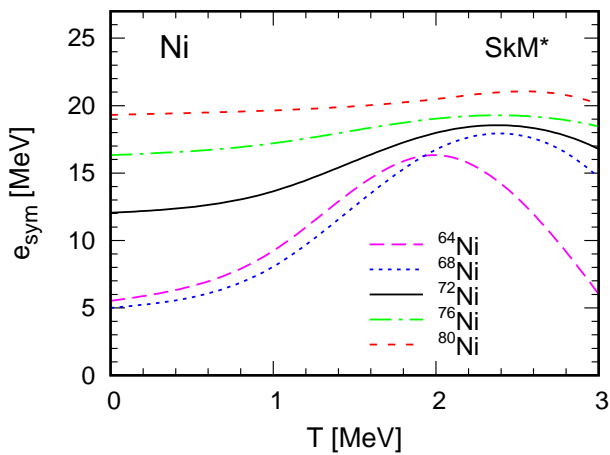}
\caption[]{(Color online) Temperature dependence of the symmetry
energy coefficient $e_{sym}$ obtained by using Eq.~(\ref{eq:8a})
for several nuclei from Ni ($A$=64--80) isotopic chain with SkM*
force. The nucleon densities and kinetic energy densities used to
calculate $e_{sym}$ are consistently derived from HFBTHO code.
\label{fig11}}
\end{figure}

\begin{figure*}
\includegraphics[width=0.45\linewidth]{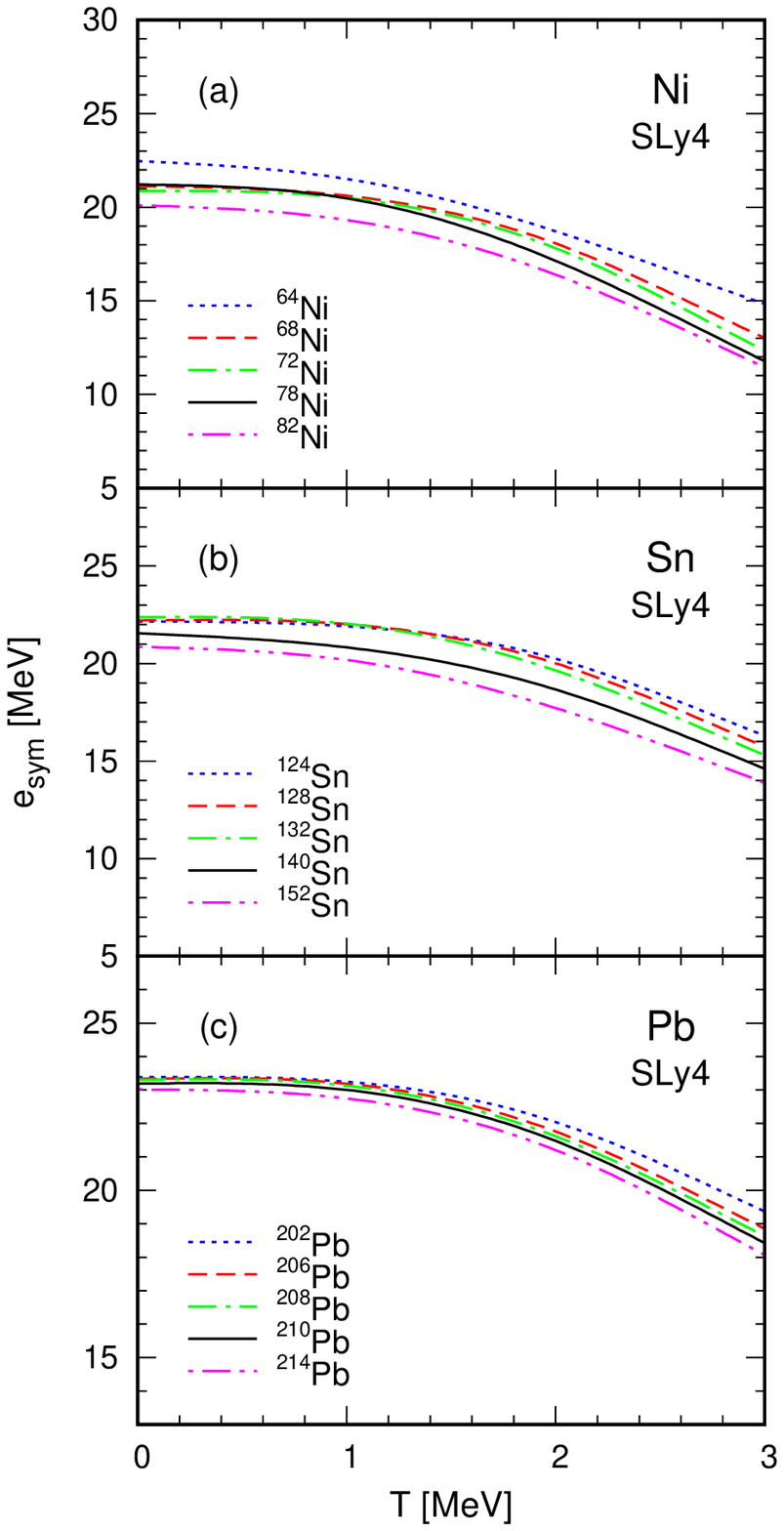}
\includegraphics[width=0.45\linewidth]{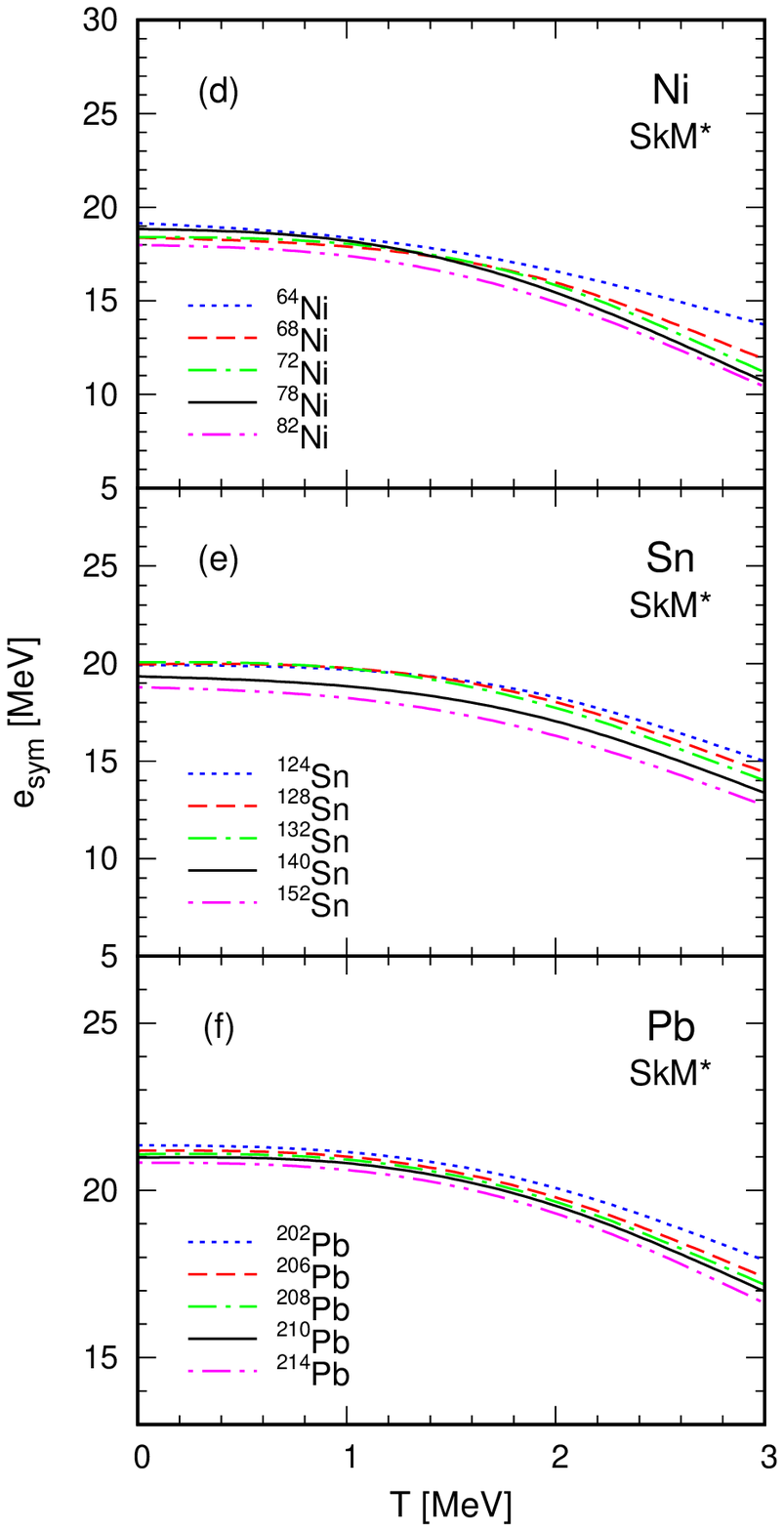}
\caption{(Color online) Temperature dependence of the symmetry
energy coefficient $e_{sym}$ obtained for several nuclei from Ni
($A$=64--82) [(a) and (d)], Sn ($A$=124--152) [(b) and (e)], and
Pb ($A$=202--214) [(c) and (f)] isotopic chains in HFB method with
SLy4 (left panel) and SkM* (right panel) forces. The results of
$e_{sym}$ are obtained by using Eqs.~(\ref{eq:1})-(\ref{eq:4})
with HFBTHO densities and $T^{2}$-approximation for the kinetic
energy density [Eq.~(\ref{eq:5})].
\label{fig12}}
\end{figure*}

As a next step of our work, we give in Fig.~\ref{fig11} the
results for the symmetry energy coefficient of five Ni isotopes
obtained by using Eq.~(\ref{eq:8a}) and SkM* force. The same
difficulties noted above at the discussion of the results
presented in Fig.~\ref{fig10} and obtained by using
Eq.~(\ref{eq:8}), appear in this case. We limited ourselves to
these cases because, as mentioned in Sec.~II~D, the even-even
nucleus with $N=Z=\bar{A}/2$ ($\bar{A}=A$) should be bound. This
is possible only for Ni isotopes but not for Sn and Pb ones. For
instance, in the case of Sn isotopes all the $N=Z$ nuclei with
$\bar{A}$ ($\bar{A}=A$) starting at 124 ($N=Z=62$) are unbound.
So, we consider the cases $^{64}$Ni: $N=Z=32$ ($^{64}$Ge),
$^{68}$Ni: $N=Z=34$ ($^{68}$Se), $^{72}$Ni: $N=Z=36$ ($^{72}$Kr),
$^{76}$Ni: $N=Z=38$ ($^{76}$Kr), $^{80}$Ni: $N=Z=40$ ($^{80}$Zr).
In contrast to the results presented in Fig.~\ref{fig10} and
further in Fig.~\ref{fig12}, the $e_{sym}(A,T)$ for the Ni
isotopes calculated using Eq.~(\ref{eq:8a}) and shown in
Fig.~\ref{fig11} do not decrease smoothly and have a different
behavior. As already mentioned above, the question of calculating
the symmetry energy coefficient for heavy nuclei with a large
isospin asymmetry needs more efforts in order to overcome the
ambiguities of the results for $e_{sym}(A,T)$ in finite nuclei
using various definitions. In our work we suggested and used two
possible ways to solve the problem. The quite different results
obtained in both cases show the strong dependence of the symmetry
energy coefficient for finite nuclei on the proper definition.

\begin{figure*}
\includegraphics[width=0.45\linewidth]{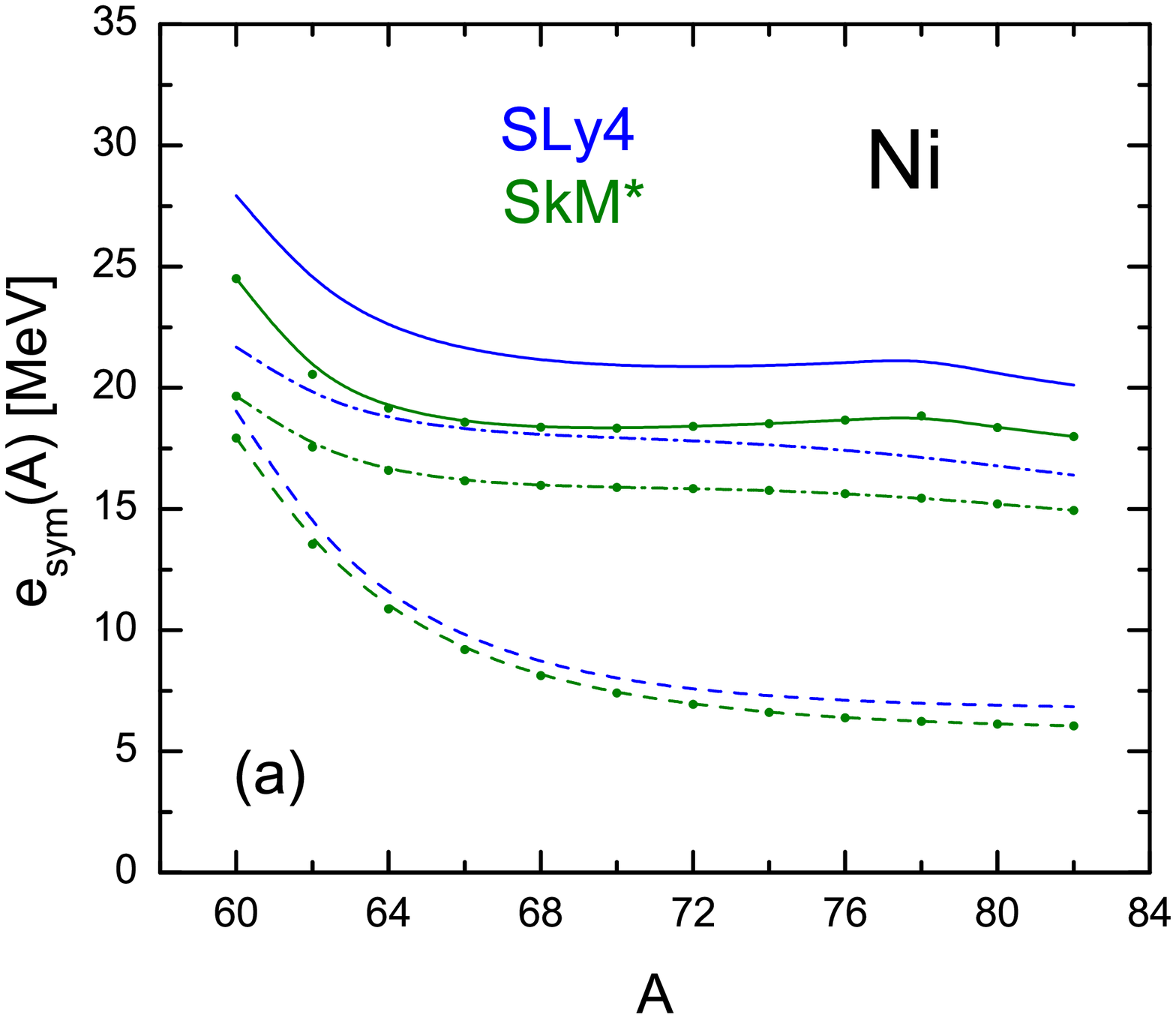}
\includegraphics[width=0.45\linewidth]{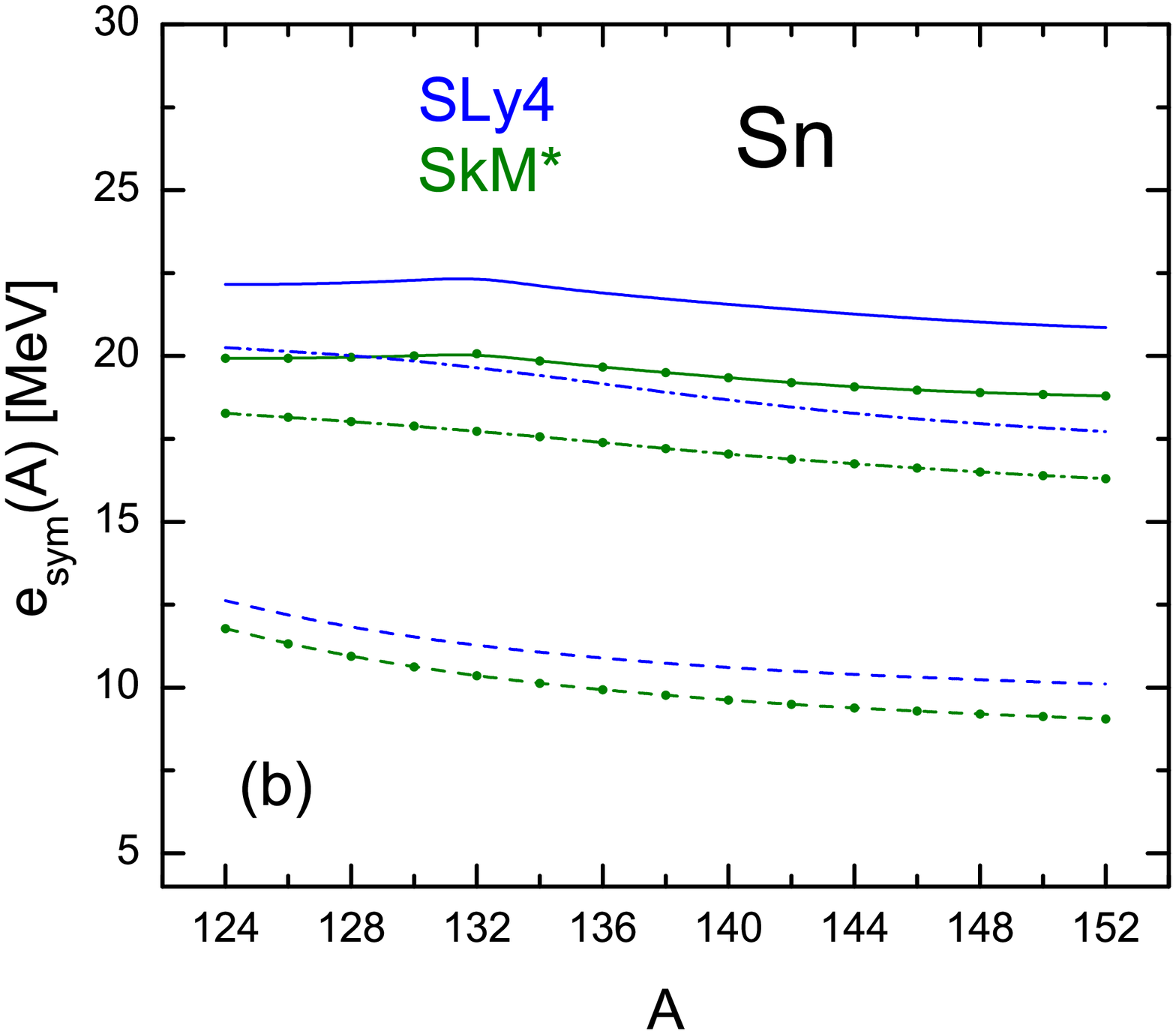}
\includegraphics[width=0.45\linewidth]{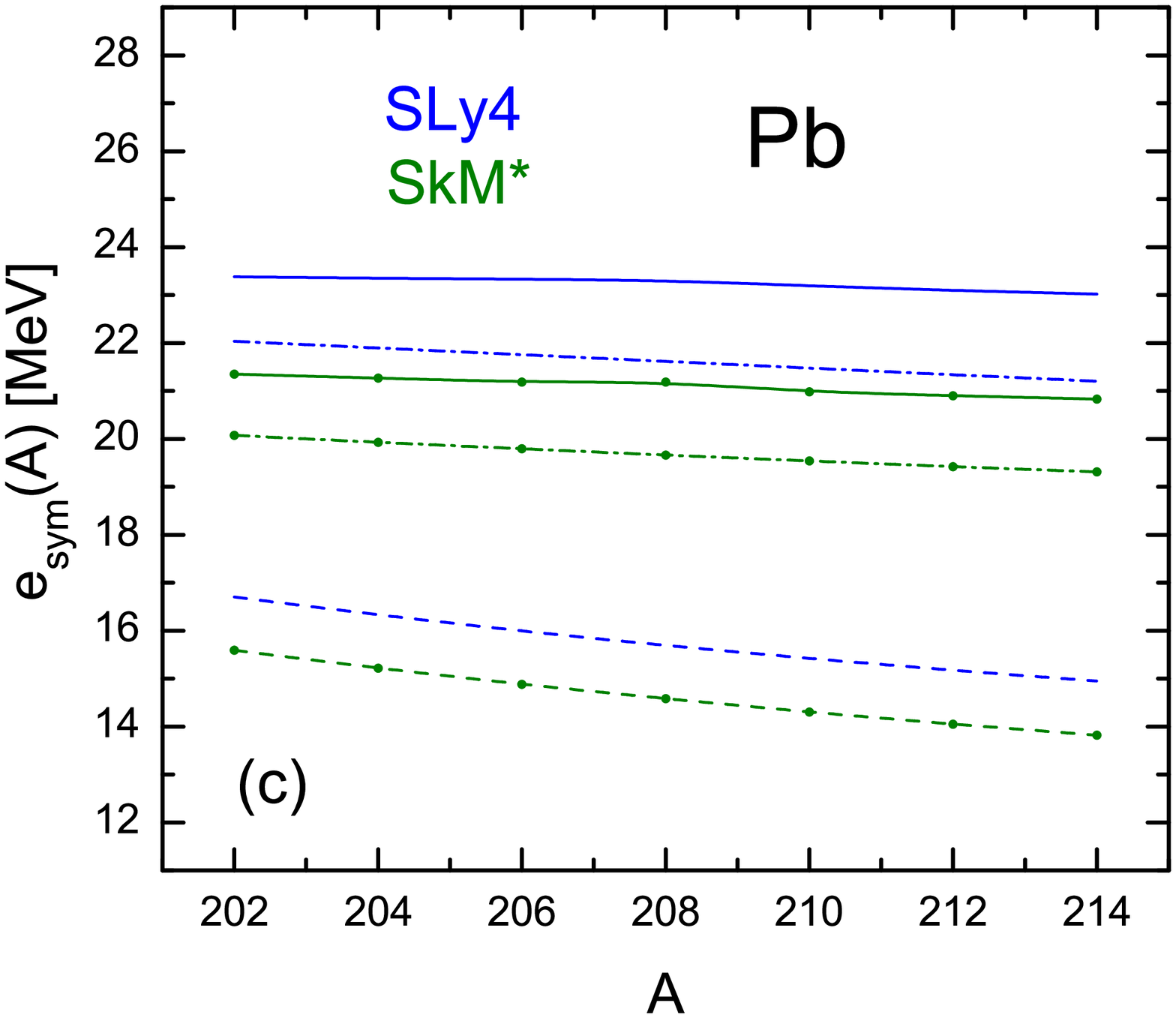}
\caption{(Color online) The mass dependence of the symmetry energy
coefficient $e_{sym}$ for Ni (a), Sn (b), and Pb (c) isotopic
chains at temperatures $T=0$ MeV (solid line), $T=2$ MeV
(dash-dotted line), and $T=4$ MeV (dashed line) calculated with
SLy4 (blue lines) and SkM* (green lines with points) Skyrme
interactions. The results of $e_{sym}$ are obtained by using
Eqs.~(\ref{eq:1})-(\ref{eq:4}) with HFBTHO densities and
$T^{2}$-approximation for the kinetic energy density
[Eq.~(\ref{eq:5})]. \label{fig13}}
\end{figure*}

For completeness, we perform a comparative analysis of $e_{sym}$
for several isotopes from the same Ni, Sn, and Pb chains applying
the LDA in a version based on Eqs.~(\ref{eq:1})-(\ref{eq:4}). The
symmetric nuclear matter part of Eq.~(\ref{eq:2})
$e(\rho,\delta=0,T)$ is obtained approximately with densities
$\rho_{n}=\rho_{p}=\rho/2$, where $\rho$ is the total density
calculated with the HFBTHO code. The kinetic energy density is
from the TF method with $T^{2}$ term \cite{Lee2010} in
Eq.~(\ref{eq:5}) calculated with the above densities. So, in this
case $\tau_{n}\approx \tau_{p}$. The results are presented in
Figs.~\ref{fig12} and \ref{fig13}. Figure \ref{fig12} illustrates
the isotopic evolution of the symmetry energy coefficient on the
examples of Ni ($A$=64--82), Sn ($A$=124--152), and Pb
($A$=202--214) chains in the case of both SLy4 and SkM* Skyrme
interactions used in the calculations. A smooth decrease of
$e_{sym}$ is observed with the increase of the mass number.
Unfortunately, it is difficult to compare our results with other
theoretical calculations of $e_{sym}$ of nuclei from the mass
range covered in the present work except from the results for
$^{208}$Pb shown in Refs.~\cite{Agrawal2014,Zhang2014} and for the
mass number $A=120$ presented in Fig.~5 of Ref.~\cite{De2012}. The
mass dependence of $e_{sym}(A)$ calculated by using the same
road-map [Eqs.~(\ref{eq:1})-(\ref{eq:5})] and densities from the
HFBTHO code is displayed in Fig.~\ref{fig13} for Ni, Sn, and Pb
isotopic chains for the same SLy4 and SkM* interactions at three
temperatures, $T$=0, 2, and 4 MeV. From one hand, one can see that
the  values of $e_{sym}$ calculated with SLy4 overestimate those
obtained with SkM* force. From another side, the difference
between both sets of values decreases going to higher
temperatures, in a way that it is small at the transition from
$T$=0 to $T$=2 MeV and a "gap" appears between the results
corresponding to $T$=2 and $T$=4 MeV. For Pb isotopic chain even a
"crossover" of curves that correspond to temperatures $T$=0 and
$T$=2 MeV and both parametrizations is observed in
Fig.~\ref{fig13}(c). We also would like to note the existence of a
kink in the values of $e_{sym}(A)$ at zero temperature at the
double-magic $^{78}$Ni and $^{132}$Sn nuclei (see
Figs.~\ref{fig13}(a) and \ref{fig13}(b)) as well as the lack of
kinks in the Pb isotopic chain [Fig.~\ref{fig13}(c)]. These
results confirm our previous observations when studying the
density dependence of the symmetry energy for Ni, Sn, and Pb
isotopes \cite{Gaidarov2011,Gaidarov2012}. We also note that in
the cases of $e_{sym}(A)$ for Ni and Sn isotopic chains the kinks
exist for $T=0$ MeV, but not for $T=2$ and $T=4$ MeV. The reason
is the well-known fact that the shell effects can be expected up
to $T\leq 2$ MeV. One can see that the values of $e_{sym}$ of
isotopes in the three chains at $T=0$ MeV obtained by following
this procedure are larger than those shown in Fig.~\ref{fig10}
obtained by using Eq.~(\ref{eq:8}) with densities and kinetic
energy densities obtained consistently using the HFBTHO code. This
can be due to the not realistic choice of the reference nucleus in
the Pb chain ($^{100}$Sn) within the previous procedure
[Eq.~(\ref{eq:8})].


\begin{figure}
\centering
\includegraphics[width=80mm]{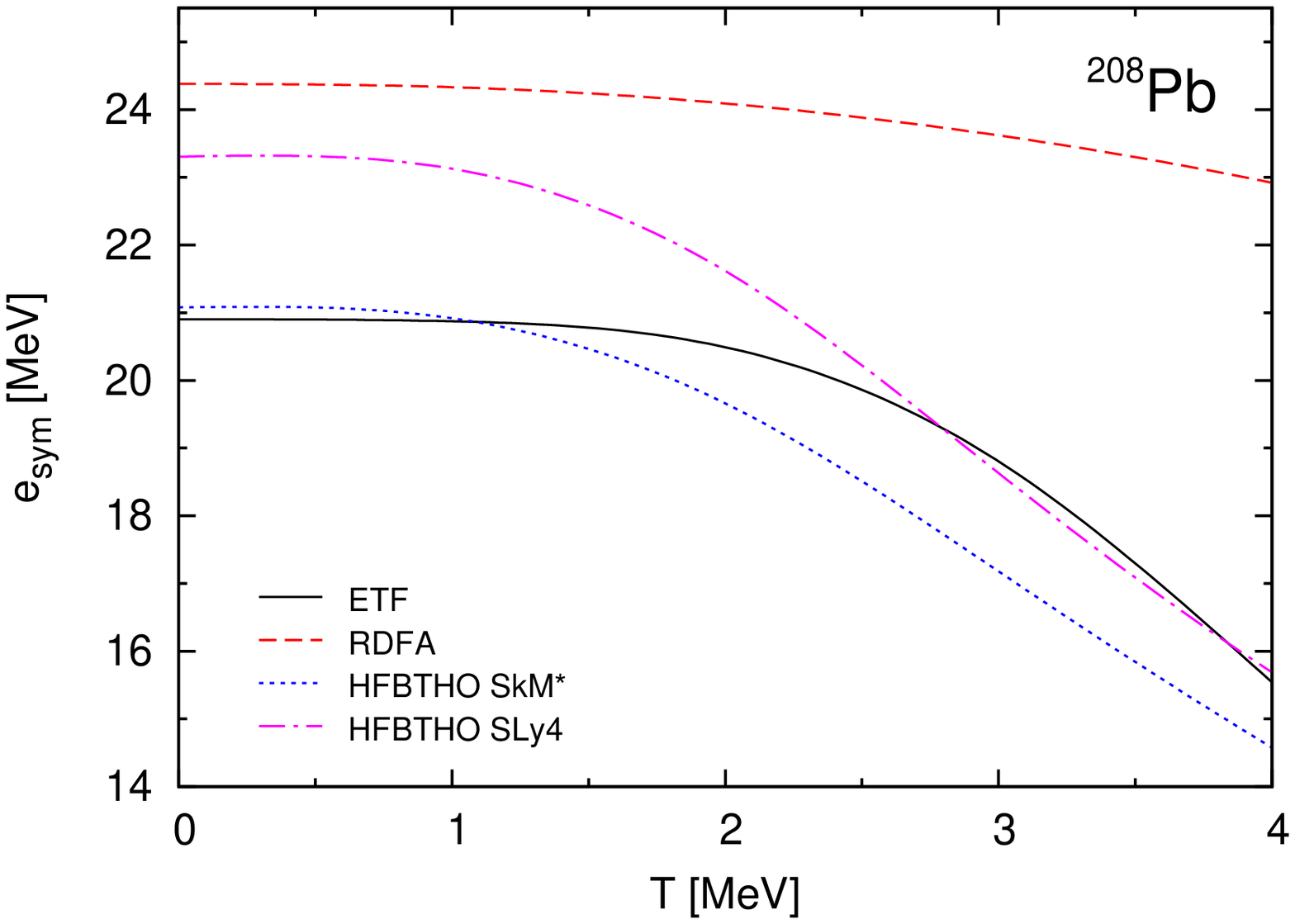}
\caption[]{(Color online) Comparison of the results for the
symmetry energy coefficient $e_{sym}$ for $^{208}$Pb calculated
with ETF [Eq.~(\ref{eq:6})], RDFA [Eq.~(\ref{eq:7})], and HFB
(with SkM* and SLy4 forces) densities. They are obtained by using
Eqs.~(\ref{eq:1})-(\ref{eq:4}) and $T^{2}$-approximation for the
kinetic energy density [Eq.~(\ref{eq:5})].
\label{fig14}}
\end{figure}

As a next step we present in Fig.~\ref{fig14} the results for
$^{208}$Pb obtained using Eqs.~(\ref{eq:1})-(\ref{eq:5}) with
three different densities, namely those obtained within the ETF,
RDFA, and HFB (with SkM* and SLy4 forces) methods. The kinetic
energy densities are obtained within TF method with $T^{2}$ term
[Eq.~(\ref{eq:5})]. The results for the thermal evolution of the
symmetry energy coefficient in the interval $T$=0--4 MeV show that
its values decrease with temperature being larger in the case of
symmetrized-Fermi density of $^{208}$Pb obtained within the RDFA.
As already discussed, the applications of different methods fail
to give unique values for the symmetry energies for finite nuclei
or their temperature dependence. Nevertheless, we would like to
note that our results for $e_{sym}$ are close to the result
obtained within the LDA (in a version reported in
Ref.~\cite{Agrawal2014}) and within the relativistic TF
approximation in Ref.~\cite{Zhang2014} for the same nucleus. The
differences in the results can be referred to the different
calculation ingredients (nucleon densities, kinetic energy density
etc.) or the adopted procedure to obtain the symmetry energy
coefficient.

\section{Conclusions}

In this work, a theoretical approach to the nuclear many-body problem has been used to study the temperature dependence of the symmetry energy coefficient in finite
nuclei and other properties, such as the $T$-dependent nucleon densities and related rms radii, as well as the possibility of formation of neutron skins. The
approach uses as a ground previous considerations within the local-density approximation (e.g., Refs.~\cite{Agrawal2014,Samaddar2007,Samaddar2008,De2012}) combining
it with the self-consistent Skyrme-HFB method using the cylindrical transformed deformed harmonic oscillator basis \cite{Stoitsov2013,Stoitsov2005}. For infinite
nuclear matter a Skyrme energy density functional with SkM* and SLy4 parametrizations is used. In our work we consider the isotopic chains of neutron-rich Ni, Sn,
and Pb isotopes that represent an interest for future measurements with radioactive exotic beams. In addition to the HFBTHO densities of these isotopes, two other
temperature-dependent densities of $^{208}$Pb were used in the present paper: the local densities within the ETF method \cite{Brack84,Brack85} that reproduce the
averaged THF results up to temperature $T$=4 MeV, and the symmetrized-Fermi local density distribution determined within the RDFA \cite{Stoitsov87}. The properties
of hot nuclei were modelled in a temperature range $T$=0--4 MeV. We have found that the ETF and RDFA results for the density distributions demonstrate a smooth
function with $r$ at any temperature $T$, while the Skyrme HFB densities have a stronger $T$-dependence. In general, the density distributions decrease with the
temperature in the center of the nucleus. Following the trend of the corresponding proton and neutron rms radii, the neutron-skin thickness grows significantly with
the increase of $T$ within a given isotopic chain. The calculated neutron-skin thicknesses by using HFBTHO densities show similar results when both SLy4 and SkM*
interactions are used. Second, we find that at zero temperature a formation of a neutron skin can be expected to start at $A>78$ and $A>132$ for Ni and Sn isotopes,
respectively, thus confirming our previously obtained results in Refs.~\cite{Sarriguren2007,Gaidarov2011}.

Our investigations of the $T$-dependent symmetry coefficients $e_{sym}(A,T)$ for finite nuclei (in particular, cases of Ni, Sn, and Pb isotopic chains) within the
LDA with some modifications face the problem for the choice of density distributions and the kinetic energy densities. In our work both quantities are calculated
through the HFBTHO code that solves the nuclear Skyrme-Hartree-Fock-Bogolyubov problem by using the cylindrical transformed deformed harmonic-oscillator basis
\cite{Stoitsov2005}. We have explored the LDA expression [Eq.~(\ref{eq:1})] for the symmetry energy $e_{sym}(A,T)$, as well as Eq.~(\ref{eq:2}), the Skyrme energy
density functional ${\cal E}(r,T)$ (the first three lines of Eq.~(\ref{eq:3})) and the nucleon effective mass [Eq.~(\ref{eq:4})]. Aiming to study the $T$-dependence
of $e_{sym}$ within a given isotopic chain, we introduced two new definitions of $e_{sym}(A,T)$ [Eqs.~(\ref{eq:8}) and (\ref{eq:8a})] within the LDA, as an attempt
to analyze in a more appropriate way the symmetry energy coefficient of finite nuclei within a given chain. Particularly, for the cases when there is no $Z=N=A/2$
bound nucleus HFB solution, none of the recipes used seems to be totaly justified or free from ambiguities, so that more work along this line is required. It is
demonstrated that using Eq.~(\ref{eq:8}), the thermal sensitivity of the symmetry energy coefficient (Fig.~\ref{fig10}) is comparatively weaker than the one
revealed when using the procedure based on Eqs.~(\ref{eq:1})-(\ref{eq:5}). In general, the results of $e_{sym}$ calculated for various isotopes in the present work
are in good agreement with theoretical predictions for some specific nuclei reported by other authors. At the same time, however, the difference between the results
given for example, in Figs.~\ref{fig10}(a) and \ref{fig11} (obtained using Eqs.~(\ref{eq:8}) and (\ref{eq:8a}), respectively) points out the dependence of the
calculations of $e_{sym}(A,T)$ on various definitions of this quantity.

Additionally, we perform a comparative analysis of $e_{sym}$ using
the procedure given by Eqs.~(\ref{eq:1})--(\ref{eq:4}), in which
the kinetic energy densities are obtained from the extension of
the TF method up to $T^{2}$ term \cite{Lee2010} [Eq.~(\ref{eq:5})]
and with HFBTHO densities. The results for the thermal evolution
of the symmetry energy coefficient of all isotopes obtained by the
procedure (\ref{eq:1})-(\ref{eq:5}) show that its values decrease
with temperature (Fig.~\ref{fig12}). This is observed also in the
particular case of $^{208}$Pb nucleus, for which different
densities have been tested to get $e_{sym}$. It is found from the
comparison (see Fig.~\ref{fig14}) that the use of
symmetrized-Fermi density obtained within the RDFA
[Eq.~(\ref{eq:7})] leads to larger values of the symmetry energy
coefficient. At the same time, for all isotopic chains considered
and for both Skyrme forces used in the calculations the symmetry
energy coefficient decreases smoothly with the increase of the
mass number in the same temperature interval (Fig.~\ref{fig12}).
In addition, it comes out that SLy4 force produces larger values
of $e_{sym}$ than the SkM* force with a fast decrease of $e_{sym}$
when $T$ increases. Studying the mass dependence of the symmetry
energy coefficient (Fig.~\ref{fig13}), we would like to note also
the existence of a kink in Ni and Sn isotopic chains at the
double-magic $^{78}$Ni and $^{132}$Sn nuclei at $T=0$ MeV,
respectively, and a lack of kink in Pb chain. This observation
confirms the result obtained previously in our works
\cite{Gaidarov2011,Gaidarov2012} when studying the nuclear
symmetry energy of spherical neutron-rich nuclei, particularly its
isotopic evolution. We pointed out that the values of $e_{sym}$ at
$T=0$ MeV obtained within this procedure for the considered three
chains are larger than those obtained by using Eq.~(\ref{eq:8})
(Fig.~\ref{fig10}) with densities and kinetic energy densities
from the HFBTHO code. For Pb isotopes the values of $e_{sym}$ are
larger than those for Ni and Sn chains.

Having in mind the dependence of $e_{sym}(A,T)$ on its various
definitions we note that more refined future investigations, for
instance, of the temperature dependence of both volume and surface
components of the symmetry energy coefficient \cite{De2012}, would
provide better description of hot nuclei and could minimize the
ambiguities due to the use of different definitions for the
symmetry energy coefficient of finite nuclei. These studies based
on our previous work \cite{Antonov2016} and the present one are in
progress.

\begin{acknowledgments}
Three of the authors (M.K.G., A.N.A., and D.N.K) are grateful for
support of the Bulgarian Science Fund under Contract
No.~DFNI-T02/19. D.N.K. thanks for the partial support from
Contract No.~DFNI-E02/6 of the Bulgarian Science Fund. E.M.G. and
P.S. acknowledge support from MINECO (Spain) under Contract
FIS2014--51971--P.
\end{acknowledgments}

\end{document}